\documentclass[preprints,article,accept,pdftex,moreauthors]{Definitions/mdpi} 
\DeclareGraphicsExtensions{.pdf,.prb,.jpg}
\usepackage[version=4]{mhchem}

\newcommand{\Lfn}[1]{{\cal L}^{(#1)}}

\newcommand{\JPB}[1]{{#1}}

\graphicspath{{./Figures_final/}}

\newcommand\ve{\varepsilon}

\firstpage{1} 
\pubvolume{1}
\issuenum{1}
\articlenumber{0}
\pubyear{2025}
\copyrightyear{2025}
\datereceived{ } 
\daterevised{ } % Comment out if no revised date
\dateaccepted{ } 
\datepublished{ } 
\hreflink{https://doi.org/} % If needed use \linebreak
\Title{Many-Body Effects in a Molecular Quantum \texttt{\texttt{NAND}} Tree}
\TitleCitation{Many-Body Effects in a Molecular Quantum \texttt{\texttt{NAND}} Tree}

\Author{Justin P. Bergfield$^{1}$\orcidA{}} %, Firstname Lastname $^{2}$ and Firstname Lastname $^{2,}$*}

\AuthorNames{Justin P. Bergfield}

\address{%
$^{1}$ \quad jpbergf@ilstu.edu; Department of Physics, Illinois State University, Normal, IL 61761, USA }

\corres{Correspondence: jpbergf@ilstu.edu}

\abstract{
Molecules provide the smallest possible circuits in which quantum interference and electron correlation can be engineered to perform logical operations, including the universal \texttt{NAND} gate.  
We investigate a chemically encoded quantum \texttt{NAND} \JPB{tree} based on alkynyl-extended iso-polyacetylene backbones, where inputs are set by end-group substitution and outputs are read from the presence or absence of transmission nodes.  
Using quantum many-body transport theory, we show that \texttt{NAND} behavior persists in the presence of dynamic correlations, but that the nodal positions and their chemical shifts depend sensitively on electron–electron interactions. This sensitivity highlights the potential of these systems not only to probe the strength of electronic correlations but also to harness them in shaping logical response. \JPB{The thermopower is identified} 
as a chemically robust readout of gate logic, providing discrimination margins that greatly exceed typical experimental uncertainties, in an observable governed primarily by the variation of transport rather than its absolute magnitude.  
}


\begin{document}

%%% C
\section{Introduction}

Quantum mechanics has transformed our understanding of nature and, through the advent of quantum algorithms, revealed a path to forms of computation that exceed the reach of classical machines.
% The rise of quantum algorithms has revealed, with striking clarity, that the principles of quantum mechanics can be harnessed to surpass the fundamental limits of classical computation.
% The development of quantum algorithms has established that quantum mechanics can be harnessed to transcend the limits of classical computational machines.
%The  advent of quantum algorithms has revealed computational advantages that transcend the capabilities of classical machines. Early 
%The development of quantum algorithms has revealed computational advantages that transcend the capabilities of classical machines. 
Early landmark results, such as the Deutsch-Jozsa algorithm for distinguishing balanced from constant functions in a single evaluation~\cite{deutsch1992rapid} and Shor’s polynomial-time algorithm for integer factorization and discrete logarithms \cite{shor1999polynomial}, demonstrate that quantum mechanical resources have the potential to circumvent specific limitations of classical architectures. Subsequent developments established quantum walks, i.e. continuous-time unitary evolutions on graphs, as a universal paradigm for quantum computation \cite{childs2009universal,childs2013universal,venegas2012quantum}, with a direct connection to scattering theory and the properties of the underlying Hamiltonians \cite{childs2011levinson,childs2012levinson,whitfield2010quantum}.

%% D
Within this framework, the \texttt{NAND} gate holds a place of particular prominence: because \texttt{NAND} is universal for classical computation, its efficient realization in a quantum setting is of fundamental conceptual interest.
%Within this framework, the \texttt{NAND} gate holds a place of prominence. Because \texttt{NAND} is universal for classical computation, its efficient evaluation in a quantum setting is of conceptual interest. 
Farhi, Goldstone, and Gutmann showed that a \texttt{NAND} \emph{tree}, \JPB{a recursive branched composition of \texttt{NAND} gates, can be evaluated via quantum scattering on a graph using only $\mathcal{O}(\sqrt{N})$ {queries}~\cite{farhi1998quantum, farhi2007quantumalgorithmhamiltoniannand,hoyer2005lower} for $N$ inputs, outperforming the best known classical algorithms, which require $\mathcal{O}(N^{0.753})$ queries~\cite{saks1986probabilistic}.} \JPB{This} advantage arises because a quantum amplitude propagates coherently through all branches of the tree, effectively interrogating many inputs in superposition, with the logical output encoded in the low-energy scattering response. Although not a universal quantum gate, the quantum \texttt{NAND} tree is an instructive model of \emph{Hamiltonian engineering} and quantum speedup \cite{farhi2007quantumalgorithmhamiltoniannand,farhi1998quantum,childs2009universal,tsuji2018quantum,  namarvar2019quantum}.

\JPB{Molecular systems, and junctions in particular, }
%Molecular junctions 
provide a natural physical system to investigate this concept \cite{baytekin2000molecular,erbas2018molecular,kompa2001molecular,evers2020advances}. When macroscopic electrodes couple to a single conjugated organic molecule, its $\pi$-system can often be represented by a tight-binding Hamiltonian in which each $p_z$ orbital defines a site and chemical bonds set the couplings. In this way, the molecular Hamiltonian mirrors the graph structure of a logic or computational problem~\cite{jensen2019molecular,soe2011manipulating,joachim2012different,renaud2008design,joachim2000electronics}. 
%Molecular junctions provide a natural physical system to investigate this concept. For example, when macroscopic electrodes couple to a single organic molecule, its $\pi$-conjugated system may often be described by a tight-binding Hamiltonian, with each $p_z$ orbital defining a site and chemical bonds setting the couplings. In this way, the molecular Hamiltonian mirrors the graph structure of a computational problem.\cite{jensen2019molecular,soe2011manipulating,joachim2012different,renaud2008design,joachim2000electronics}
%Molecular junctions offer a natural physical system with which to investigate this concept. In such systems, macroscopic electrodes couple to a single molecule, which can be chosen such that its $\pi$-conjugated system is well described by a tight-binding Hamiltonian. Each $p_z$ orbital defines a site, and chemical bonds set the couplings, so that the molecular Hamiltonian mirrors the graph structure of a computational problem.\cite{jensen2019molecular, soe2011manipulating, joachim2012different, renaud2008design, joachim2000electronics} 
In particular, cross-conjugated scaffolds recreate the interference motifs of the \texttt{NAND} tree: substituents at branch termini act as controlled self-energies that shift, create, or annihilate interference nodes. Inputs are thereby encoded as chemical modifications, while the logical output is read from whether a transmission node occurs at the Fermi energy (\textsc{off}) or is shifted so as to permit conduction (\textsc{on}). This mapping is the essence of molecular Hamiltonian computing~\cite{dridi2015mathematics,dridi2018qubits,joachim2005hamiltonian,joachim2012different}.

%At the heart of these approaches lie transmission nodes, spectral features created through destructive quantum interference (QI) in which all transport amplitudes cancel, encoding the underlying symmetries and connectivities of the Hamiltonian.\cite{bergfield2011novel,barr2013transmission,pedersen2014quantum}
%At the heart of these realizations lie transmission nodes, spectral features stemming from destructive quantum interference (QI) whereby all transport amplitudes  cancel exactly, effectively encoding fundamental  symmetries and connectivities of the Hamiltonian.\cite{bergfield2011novel,barr2013transmission,pedersen2014quantum} 
%Crucially, such nodes have been observed directly in single-molecule junctions,\cite{guedon2012observation,li2019gate,liu2018quantum,nairz2003quantum} even at room temperature, underscoring the robustness of coherent transport in these systems. Because transport through SMJs is largely elastic and phase coherent due to strong charging energies,\cite{aradhya2012dissecting,arroyo2013signatures,wang2020thermal} they provide a tangible bridge between abstract scattering-based computation and chemically realizable logic elements.

At the heart of these approaches lie transmission nodes, spectral features created through destructive quantum interference (QI) in which all transport amplitudes cancel, encoding the underlying symmetries and connectivities of the Hamiltonian.\cite{bergfield2011novel,barr2013transmission,pedersen2014quantum,arroyo2013signatures} Such nodes have been observed in single-molecule junctions, even at room temperature, underscoring the robustness of coherent transport in these systems~\cite{guedon2012observation,li2019gate,liu2018quantum,nairz2003quantum,aradhya2012dissecting,wang2020scale,liu2023signatures}. \JPB{Although this work focuses on theory, experimental references are included solely to provide context and motivation. Because transport through small single-molecule junctions (SMJs) is predominently elastic and quantum phase coherent, they provide a natural bridge between abstract scattering-based computation and chemically realizable logic elements.} 

Importantly, QI features also leave a distinct imprint on the thermoelectric response~\cite{bergfield2009many,bergfield2010giant,bennett2024quantum,erdogan2025dephasingfailsthermodynamicconsequences,miao2018influence,bergfield2024identifying, hamill2023quantum, Finch09}. As a transmission node is approached, charge and entropy currents vanish at different rates; their ratio, the Seebeck coefficient, can be strongly enhanced, serving both as a potential route to improved thermoelectric performance and as a practical diagnostic of interference zeros.
Whereas conductance is often highly sensitive to variations in contact hybridization, molecular conformation, and other microscopic details~\cite{ke2008quantum,cardamone2006controlling, hansen2009interfering, markussen2010relation,Solomon08,solomon2009electron} which may obscure the readout of a quantum circuit, the thermopower reflects the first transport moment of the transmission function via the Mott relation \cite{lunde2005mott}. 
\JPB{
As such, the thermopower for energies near the node is largely insensitive to the microscopic junction details when the nodes is detuned from resonance~\cite{bennett2024quantum}, but remains highly sensitive to the position, scaling, and dephasing of those nodes \cite{bergfield2009thermoelectric, bergfield2025quantuminterferencesupernodesthermoelectric,inui2018emergence}, making it a robust indicator of nodal structure and, by extension, of the logical state.}

A central limitation of many proposals, however, is their reliance on independent-electron tight-binding %(TB) 
descriptions. While these methods capture the role of junction topology, they neglect dynamical electron-electron correlations which dominate transport at the nanoscale. In acyclic cross-conjugated polymers, for instance \cite{pedersen2014quantum,barr2013transmission,solomon2011small}, long-range Coulomb interactions couple distant orbitals, modifying the phases of interfering amplitudes and shifting nodal positions. Because these interactions decay only slowly with distance, the Hamiltonian requires increasingly delicate fine tuning to reproduce a digital \texttt{NAND}-like response, and the tuning becomes more fragile precisely in the regime where interference is most pronounced.

In this article, a state-of-the-art many-body theory is used to investigate the steady-state transport through molecular implementations of the \texttt{NAND} tree, constructed from alkynyl-extended iso-polyacetylene backbones \cite{bergfield2009many}. Logical inputs are encoded chemically through \ce{NH2} or vinyl substitution, while outputs are read from the presence or absence of a transmission node in the junction \cite{jensen2019molecular}. The alkynyl linkers suppress extraneous $\sigma$-channels, isolating the $\pi$-interference that governs low-energy transport.
We find that both independent-electron (H\"uckel) and many-body descriptions reproduce \texttt{NAND} gate behavior, demonstrating that the logical motif survives dynamic electron–electron correlations. However, they predict different nodal energies and opposite  substitution trends, differences that turn these junctions into natural \emph{diagnostics} of correlation strength and expose the limits of simplified ``stub-resonator'' models.
We also investigate the use of thermopower as a means to read the logical state, finding that it is insensitive to chemical details and retains clear discriminatory power even when conductance signatures blur \cite{bennett2024quantum,bergfield2024identifying}.
Beyond supporting the original molecular \texttt{NAND}-gate proposals, our results extend the concept into the many-body Fock space, where correlations can reshape, and even generate, nodes which have no single-particle analog \cite{bergfield2011novel}.

\section{From \texttt{NAND} trees to quantum transport}

%% B
\JPB{
Motivated by the topological similarity between certain acyclic cross-conjugated molecules and \texttt{NAND} trees, we now develop a minimal {transport formulation} that makes this mapping precise. In the Farhi–Goldstone–Gutmann construction \cite{farhi2007quantumalgorithmhamiltoniannand}, the NAND tree is evaluated as a \emph{scattering problem}, where the logical output is encoded in the transmission or reflection amplitude of a particle propagating coherently through the connected tree. This approach naturally connects with quantum transport theory~\cite{meir1992landauer,imry1999conductance}, where interference effects like nodes are encoded in transmission function spectrum.  Within this framework, it is the coherent combination of amplitudes along the connected tree Hamiltonian that yields the NAND truth table and underlies the ${\cal O} (\sqrt{N})$ quantum query complexity \cite{jensen2019molecular}. It should be noted that the speed-up of this method does not stem from the use of molecules, the inclusion or exclusion of electron-electron interactions, etc., but instead from the use of quantum coherent transport to implement gate logic.
}
%% B

% We emphasize that the NAND-tree algorithm of Farhi, Goldstone, and Gutmann is a scattering algorithm. The logical output is encoded in the transmission or reflection amplitude of a particle propagating coherently through the coupled tree. Coherent tunneling between branches is therefore essential: a model without tunnel couplings, even if it includes electron–electron repulsion or onsite energy shifts, cannot yield a NAND output in transmission nor reproduce the $\tilde O(\sqrt{N})$ query complexity. The quantum advantage is a direct consequence of interference in the connected tree Hamiltonian, not of static level shifts.

% Motivated by the topological similarity between certain acyclic cross-conjugated molecules and \texttt{NAND} trees, we now develop a minimal transport formulation that makes this mapping precise. 

Using nonequilibrium Green’s function (NEGF) theory~\cite{bergfield2009many,HaugAndJauhoBook,stefanucci2013nonequilibrium,cuniberti2005introducing}, the transmission probability is given by
\begin{equation}
{\cal T}(E) = {\rm Tr}\!\left\{ \Gamma_L(E)\,{\cal G}(E)\,\Gamma_R(E)\,{\cal G}^\dag(E) \right\},
\label{eq:transmission_prob}
\end{equation}
where ${\cal G}(E)$ is the retarded Green’s function of the junction. The tunneling-width matrix for electrode $\alpha$ is
\begin{equation}
\big[\Gamma_\alpha(E)\big]_{nm} = 2\pi \sum_{k\in\alpha} V_{nk} V_{mk}^\ast\, \delta(E-\epsilon_k),
\end{equation}
with $n,m$ labeling $\pi$-orbitals of the molecule and $V_{nk}$ the coupling between orbital $n$ and eigenstate $\epsilon_k$ in electrode $\alpha$. In the wide-band limit, considered here, $\Gamma_\alpha$ may be taken as energy-independent.

%We 
\JPB{Consider a} two-terminal junction where left ($L$) and right ($R$) electrodes couple to a molecular backbone, i.e.\ a linear ``direct'' transport channel, with a locally attached side branch (stub), as illustrated schematically in Fig.~\ref{fig:NAND_schematic}. Using Dyson's equation, this junction's Green’s function may be expressed as
\begin{equation}
{\cal G} (E) = \left[ {\cal G}_{\rm back}^{-1}(E) - \Sigma_{\rm stub}(E) \right]^{-1},
\end{equation}
where the stub's influence on the backbone is encapsulated in the self-energy $\Sigma_{\rm stub}$.
If the backbone is described by a tight-binding Hamiltonian $H_{\mathrm{back}}$, its retarded Green’s function is
\begin{equation}
{\cal G}_{\mathrm{back}}(E) = \left[(E+i0^+)\mathbf{1}
- \hat{H}_{\mathrm{back}} - \Sigma_{\rm T}(E)\right]^{-1},
\end{equation}
where $\Sigma_{\rm T}(E)=\Sigma_{\rm L}(E)+\Sigma_{\rm R}(E)$ is the tunneling self-energy. Each $\Sigma_\alpha$ may be written as
$\Sigma_\alpha(E) = \Lambda_\alpha(E) - \tfrac{i}{2}\,\Gamma_\alpha(E)$,
with $\Gamma_\alpha$ the level broadening from coupling to lead $\alpha$ and $\Lambda_\alpha$ the corresponding level shift. In the wide-band limit $\Gamma_\alpha$ is taken constant and $\Lambda_\alpha$ absorbed into the onsite energies, leaving ${\rm Im}\,\Sigma_\alpha=-\Gamma_\alpha/2$. In this formulation, the stub's molecular structure, and therefore its self-energy, directly affects the observable transport, the conductance and thermopower are functions of ${\cal T}(E)$, motivating a search for stub structures that mimic logic gates.

To illustrate how such branch structures can emulate a logic gate, consider a stub consisting of a uniform chain of length $n$ with site energy $\epsilon_b$ and hopping $t_b$. The stub self-energy can be determined diagrammatically, giving
\begin{equation}
\Sigma_{\rm stub}(E) = V^\dagger g_{\rm stub}(E)V = t_b^2 g_{\rm stub}(E),
\end{equation}
where $g_{\rm stub}(E)$ is the Green’s function of the isolated $n$-site chain. This recursive structure of the Green’s function admits the continued fraction
\begin{equation}
g_{\rm stub}(E) =
\frac{1}{E-\epsilon_b-\dfrac{t_b^2}{E-\epsilon_b-
\dfrac{t_b^2}{\ddots-\dfrac{t_b^2}{E-\epsilon_b}}}}
\qquad (n\ \text{sites}),
\end{equation}
and in closed form
\begin{equation}
\Sigma_{\rm stub}(E) =
t_b\,\frac{U_{n-1}(z)}{U_n(z)},
\qquad z=\frac{E-\epsilon_b}{2t_b},
\label{eq:Sigma-Cheb}
\end{equation}
with $U_m$ representing the Chebyshev polynomials of the second kind. At resonance $E=\epsilon_b$,
\begin{equation}
\Sigma_{\rm stub}(\epsilon_b) =
\begin{cases}
0, & n\ \text{even},\\[4pt]
\infty, & n\ \text{odd},
\end{cases}
\label{eq:parity-rule}
\end{equation}
since $U_{2k}(0)=(-1)^k$ and $U_{2k+1}(0)=0$. 
% Thus an even-length branch is transparent, whereas an odd-length branch enforces a transmission node. Therefore, a side chain realizes an energy-dependent self-energy whose poles and zeros toggle conduction. Although no closed-form expression is known for the self-energy of an entire \texttt{NAND} tree,\cite{brisker2008controlled} each additional branch level simply extends the same recursion—the algebraic core of \texttt{NAND}-tree evaluation.
Thus, an even-length branch is transparent, while an odd-length branch enforces a transmission node. This  illustrates how side-chains can induce poles and zeros in the effective self-energy, thereby switching conduction on or off. While the closed-form self-energy of a \texttt{NAND} tree is not known~\cite{brisker2008controlled}, it is precisely this recursive structure that echoes the evaluation of a \texttt{NAND} tree.

Consider an SMJ in which the left ($L$) and right ($R$) electrodes couple only to a linear backbone, while a side branch (stub) is grafted locally and not directly contacted. When all current must traverse the backbone, $\Sigma_{\rm stub}\!\to\!\infty$ forces ${\cal G}_{LR}\!\to\!0$ and thus ${\cal T}\!\to\!0$ (logical \textsc{off}), whereas $\Sigma_{\rm stub}=0$ leaves the backbone unperturbed and transmission intact (logical \textsc{on}). Shifting the branch resonance, $\epsilon_b\!\to\!\epsilon_b+\delta$, continuously tunes $\Sigma_{\rm stub}(E)$ between these limits—a form of remote gating~\cite{brisker2008controlled}. More elaborate branch hierarchies can be integrated recursively, each level introducing additional poles and zeros (resonances and antiresonances). 
%
%The transport recursion mirrors \texttt{NAND}-tree evaluation but also reveals its fragility: the finite molecular bandwidth and nonuniform couplings only allow a limited number of resolvable nodes before the spectrum fragments.
%
%
This leads to a central design principle: \emph{molecular logic arises from the competition between direct backbone conduction and indirect branch-induced self-energies}. For simple motifs the mapping onto Boolean truth tables is robust, but as complexity grows, multiple resonances compete and logical discrimination becomes ambiguous.

We have thus far focused exclusively on noninteracting electrons. While the recursive self-energies of stubs \JPB{can} reproduce \texttt{NAND}-tree logic in this limit, the Coulomb self-energy $\Sigma_{\rm C}$ of an interacting junction \JPB{may} admit no closed form: it is strongly energy dependent, nonlocal, and encodes the full many-body dynamical excitation spectrum~\cite{Baym62,PhysRev.124.41,kozik2015nonexistence,vuvcivcevic2018practical}. Moreover, the Hilbert spaces themselves differ. The model above acts in a one-electron orbital basis, whereas the interacting problem requires the full many-body Fock space, which includes states and correlations beyond any single-particle description. These effects shift and split resonances, alter scattering phases, and can generate interference features unrelated to molecular geometry~\cite{bergfield2011novel}. We therefore turn next to a fully many-body theory of quantum transport, in which the molecular Green’s function is constructed nonperturbatively to investigate the importance of correlations in these systems.
%and the analytic ambiguities of skeleton expansions are avoided.\cite{bergfield2009many}

\section{Many-body Quantum Transport Theory}

% To capture electron–electron correlations beyond independent-electron models, we employ the formally exact many-body molecular Dyson equation (MDE) theory.\cite{bergfield2009many} 
Within the molecular Dyson equation (MDE) theory~\cite{bergfield2009many}, the junction Green’s function may be formally written exactly as
\begin{equation}
{\cal G}(E) = \left[ {\cal G}_{\rm mol}^{-1}(E) - \Sigma_{\rm T}(E) - \Delta \Sigma_{\rm C}(E) \right]^{-1},
\label{eq:Dyson2}
\end{equation}
where ${\cal G}_{\rm mol}$ is the isolated molecular Green’s function, including both one- and two-body (i.e. Coulomb) terms, and $\Sigma_{\rm T}=\Sigma_{\rm L} + \Sigma_{\rm R}$ is the tunneling self-energy with $\Sigma_\alpha=-i\Gamma_\alpha/2$ in the wide-band limit considered here. In this theory, the {\em correction} to the Coulomb self-energy, $\Delta\Sigma_{\rm C}$, accounts for the renormalization of the electron–electron interactions with finite lead–molecule coupling. In this work we focus on transport in the elastic cotunneling regime, where $\Delta\Sigma_{\rm C}\approx 0$ and inelastic processes are negligible~\cite{bergfield2009many}.

The molecular Green’s function admits the usual Lehmann representation,
% \begin{equation}
% {\cal G}_{\rm mol}(E) = \sum_{\Psi,\Psi'} 
% \frac{\big[{\cal P}(\Psi) + {\cal P}(\Psi')\big]\,
% C^{\Psi \rightarrow \Psi'}}{E-E_{\Psi'}+E_{\Psi}+i0^+},
% \label{eq:Gmol}
% \end{equation}
\begin{equation}
\big[{\cal G}_{\rm mol}(E)\big]_{n\sigma,m\sigma'} =
\sum_{\Psi,\Psi'} 
\frac{\big[{\cal P}(\Psi) + {\cal P}(\Psi')\big]\,
C^{\Psi \rightarrow \Psi'}_{n\sigma,m\sigma'}}{E - E_{\Psi'} + E_{\Psi} + i0^+},
\label{eq:Gmol}
\end{equation}
where $E_\Psi$ is the eigenenergy of many-body state $\left| \Psi \right.\rangle$ of $\hat{H}_{\rm mol}$, \JPB{the molecular Hamiltonian,} ${\cal P}(\Psi)$ its occupation probability, and
\begin{equation}
C^{\Psi \rightarrow \Psi'}_{n\sigma,m\sigma'} =
\langle \Psi | \hat{d}_{n\sigma} | \Psi' \rangle
\langle \Psi' | \hat{d}^\dagger_{m\sigma'} | \Psi \rangle
\label{eq:manybody_element}
\end{equation}
are the transition matrix elements. Here $\hat{d}_{n\sigma}$ annihilates an electron of spin $\sigma$ on orbital $n$, and $\left| \Psi \right.\rangle$, $\left| \Psi' \right.\rangle$ %$\Psi,\Psi'$ 
belong to the $N$- and $(N+1)$-particle \JPB{spaces}, respectively. In linear response, the probabilities ${\cal P}(\Psi)$ are given by the grand canonical ensemble.

We focus on transport in the linear-response regime, where the electrical conductance and thermopower are expressed in terms of the Onsager functions ${\cal L}^{(\nu)}$ as
\begin{align}
    G &= e^2 {\cal L}^{(0)}, \\
    S &= -\frac{1}{eT_0}\,\frac{{\cal L}^{(1)}}{{\cal L}^{(0)}},
\end{align}
with $e$ being the elementary charge, $G$ the conductance, $S$ the Seebeck coefficient, and $T_0$ the reference temperature. In the SMJs considered here, transport is overwhelmingly coherent and elastic. In this limit the Onsager functions may be expressed as
\begin{equation}\label{eq:Lfun}
{\cal L}^{(\nu)} = \frac{1}{h} \int dE \,\big(E - \mu_0 \big)^\nu 
{\cal T}(E)\left(-\frac{\partial f_0}{\partial E} \right),
\end{equation}
where ${\cal T}(E)$ is the transmission function and  $f_0(E) = \left[\exp\!\big((E-\mu_0)/k_BT_0\big)+1\right]^{-1}$
% \begin{equation}
% f_0(E) = \left[\exp\!\left(\frac{E-\mu_0}{k_BT_0}\right)+1\right]^{-1}
% \end{equation}
is the Fermi–Dirac distribution of the electrodes with equilibrium chemical potential $\mu_0$ and temperature $T_0$.
%is the Fermi–Dirac distribution describing the electrodes with equilibrium chemical potential $\mu_0$ and temperature $T_0$.

\subsection{Lanczos method for Green’s functions}

The central computational difficulty in evaluating Eq.~\eqref{eq:Gmol} is that it 
requires the full many-body spectrum, which grows exponentially with the number of 
orbitals ($\sim$4$^n$ states for $n$ orbitals). To render the problem tractable, we 
employ Krylov-space techniques, most notably Lanczos recursion~\cite{lanczos1950iteration},
which iteratively projects the Hamiltonian onto a reduced subspace. This procedure yields~\cite{barr2012effective} 
a tridiagonal representation of $\hat{H}_{\rm mol}$ and a continued-fraction expansion of 
${\cal G}(E)$ that converges rapidly for the low-energy states and spectral weights of 
interest, while avoiding explicit diagonalization~\cite{mori1965continued}.

In practice the calculation proceeds in two stages~\cite{barr2012effective}. First, a Lanczos diagonalization of 
$\hat{H}_{\rm mol}$ provides the low-lying eigenstates and their energies. Second, the action of 
$\hat{d}_m$ or $\hat{d}^\dagger_n$ on these states is used to seed a new Lanczos recursion, generating 
Krylov subspaces of virtual excitations that enter the propagator. The resulting spectral 
representation can then be evaluated directly or through the continued fraction, which by 
construction preserves causality. 
This two-stage approach enables the calculation of Green’s functions for realistic 
multi-orbital $\pi$-systems such as iso-PA backbones. In this way, many-body correlations 
are treated exactly within the truncated $\pi$-subspace, providing the rigorous foundation 
for the results presented below.

%% C
\subsection{Molecular Hamiltonian}

%The tranpsort through 
\JPB{This work focuses on the response of acyclic cross-conjugated molecules, where conduction is dominated by the $\pi$-system.}
%We focus on the response of acyclic cross-conjugated molecules, where conduction is dominated by the $\pi$-system. 
The Hamiltonian for this subspace was derived using a renormalization procedure that integrates out off-resonant degrees of freedom, such as the $\sigma$-system and image-charge effects, so that their influence appears effectively through modified onsite energies and coupling terms~\cite{barr2012effective}. In a localized orbital basis, the Hamiltonian reads
\begin{align}
\label{eq:Hmol}
\hat{H}_{\rm mol} & = \sum_{n,\sigma} \ve_{n\sigma} \hat{\rho}_{n\sigma}
- \sum_{\langle n,m \rangle,\sigma} t_{nm} \hat{d}^\dagger_{n\sigma} \hat{d}_{m\sigma} \nonumber 
+ \tfrac{1}{2} \sum_{nm} U_{nm} \hat{q}_n \hat{q}_m ,
\end{align}
where $\ve_{n\sigma}$ is the effective onsite potential for spin-$\sigma$ electrons on orbital $n$, 
$\hat{\rho}_{n\sigma} = \hat{d}^\dagger_{n\sigma} \hat{d}_{n\sigma}$, 
$\hat{q}_{n} = (\sum_\sigma \hat{\rho}_{n\sigma})-1$ is the net charge operator, and $t_{nm}$ are the effective tight-binding matrix elements. The Coulomb interaction $U_{nm}$ between electrons in orbitals $n$ and $m$ is obtained from a multipole expansion, including monopole–monopole, quadrupole–monopole, and quadrupole–quadrupole contributions~\cite{barr2012effective}:
\begin{align}
U_{nm} &= U_{nn}\,\delta_{nm} + (1-\delta_{nm})\!\left(U^{MM}_{nm} + U^{QM}_{nm} + U^{MQ}_{nm} + U^{QQ}_{nm}\right).
\end{align}
The $\pi$-EFT parameters were determined by fitting to experimental observables that must be faithfully reproduced within a $\pi$-only model~\cite{barr2012effective}. Specifically, the vertical ionization energy, vertical electron affinity, and the six lowest singlet and triplet excitations of gas-phase benzene were optimized simultaneously. This procedure yields accuracy comparable to, or better than, traditional Pariser-Parr-Pople (PPP) models~\cite{barford2005electronic}, giving $U_{nn}$=9.69eV for onsite repulsion; transfer integrals $t=2.2$, $2.64$, and $3.0$~eV for single, double, and triple carbon–carbon bonds, respectively~\cite{purcell1967brief}, and a $\pi$-electron quadrupole moment $Q=-0.65\,e$\AA$^2$. These parameters are consistent with earlier $\pi$-electron models~\cite{barford2005electronic,ohno1964some}, with $Q$ providing a physically motivated alternative to the ad hoc short-range corrections of PPP theory. 
Electron–electron interactions were screened by an effective dielectric constant $\varepsilon=1.56$. Electrodes were modeled as metallic spheres of radius 0.5~nm. To allow direct comparison with prior work~\cite{jensen2019molecular}, thiol groups were not treated explicitly as sites, but incorporated via renormalized lead–molecule couplings. 
%Finally, all molecular geometries used here were optimized including all necessary thiol and hydrogen atoms required to ensure chemical stability using the MMFF94s force field,\cite{halgren1996merck} with gradient descent to obtain reliable structures for conjugated organics at modest computational cost.
% Finally, all molecular geometries were optimized with the MMFF94s force field~\cite{halgren1996merck}, with thiol and hydrogen atoms included for chemical stability and gradient descent used to obtain reliable conjugated-organic structures at modest computational cost. \JPB{Do Orca here?} %\JPB{The conclusions of this work are not critically sensitive on }
% %% C
%
\JPB{Finally, all molecular geometries were optimized using Kohn–Sham density functional theory (DFT) in ORCA 6.1.0, employing the B3LYP functional with D3(BJ) dispersion corrections and the 6-311G(d,p) (6-311G**) basis set~\cite{Becke1993,LYP1988,Krishnan1980,McLeanChandler1980,RN269,RN33,RN75,RN114,RN218}. Tight self-consistent field (SCF) thresholds were used to ensure reliable conjugated-organic structures at modest computational cost.}

\begin{figure}[tb]
	\centering
	\includegraphics[width=\linewidth]{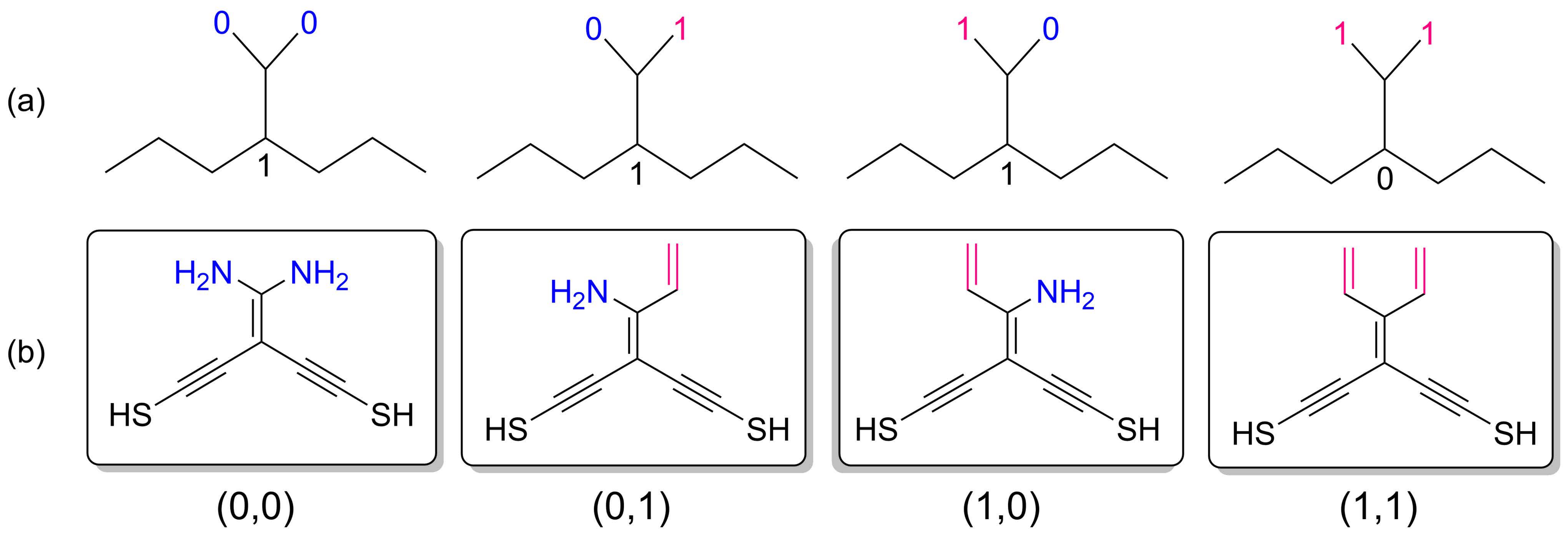}
	% \caption{Schematic diagrams of the thiolated, alkynyl-extended iso-poly(acetylene) (top) and iso-poly(diacetylene) (bottom) analogues, denoted for brevity as NCC and NCCA, respectively, in the text. Each cross-conjugated repeat unit contains terminal ``stub'' \ce{C-C} units with a terminal substituents $R$; in principle, these groups may vary and can be used to stabilize the backbones or tune onsite energies. In the present work, we take $R=\text{H}$ to reduce computational complexity and isolate the intrinsic interference features.
	% }
    % \caption{Schematic diagram (a) of a (? one level??) \texttt{NAND} tree stub with all four possible inputs and outputs at each level connected to a and (b) the molecular implementations using a a thiolated, alkynyl-extended iso-poly(acetylene) molecule with \ce{NH2} and vinyl ``inputs'' representing logical 0 and 1's, respectively. 
    % }
    % \caption{{\bf (a)} Schematic representation of a single-level \texttt{NAND} stub, illustrating the four possible input combinations and their logical outputs. Inputs are encoded as branches labeled 0 (blue) or 1 (magenta), with the \texttt{NAND} truth table reproduced by the transmission response at the nodal energy. {\bf (b)} Molecular implementations based on thiolated, alkynyl-extended iso-poly(acetylene) (iso-PA) backbones. Logical $0$ is represented by an \ce{NH2} substituent, while logical $1$ is represented by a vinyl (\ce{C=C}) group. Outputs are read from the transport, for example, when macrocopic Au electroded coupled to the thiol groups.}
    \caption{{\bf (a)} Schematic representation of a single-level \texttt{NAND} stub, illustrating the four possible input combinations and their logical outputs. Inputs are encoded as branches labeled 0 (blue) or 1 (magenta), with the \texttt{NAND} truth table reproduced by the transmission response at the nodal energy. (b) Molecular realizations based on thiolated, alkynyl-extended iso-poly(acetylene) (iso-PA) backbones. Logical $0$ is represented by an \ce{NH2} substituent, while logical $1$ is represented by a vinyl (\ce{C=C}) group. Outputs are read from the transmission spectrum near the nodal energy, as obtained in transport measurements when macroscopic Au electrodes are coupled to the terminal thiol groups.}
	\label{fig:NAND_schematic}
\end{figure}

%C:\SynologyDrive\!Research\2025_0829-Molecular_Circuit\plot_NAND_v2.m
\begin{figure}[tb]
	\centering
    %\isPreprints{}{% This command is only used for ``preprints''.
% \begin{adjustwidth}{-\extralength}{0cm}
	\begin{minipage}{.49\textwidth}
		\includegraphics[width=\linewidth]{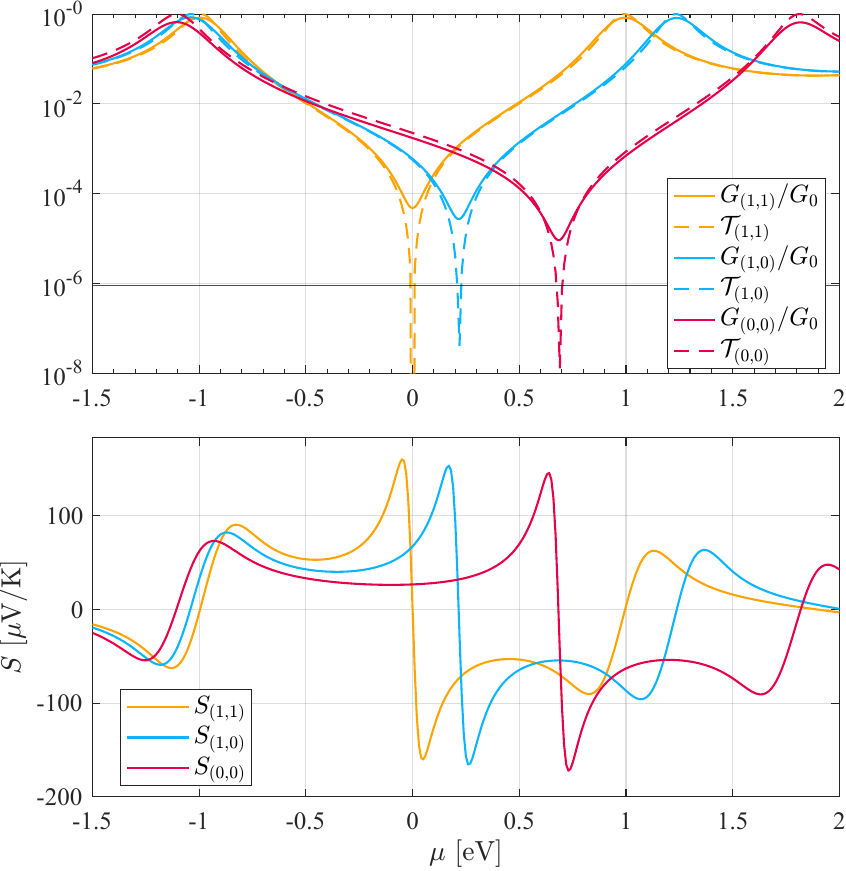}	
		\caption*{{\bf (a)} Single-particle tight-binding theory}
		%		\vspace{.5em}
	\end{minipage}
	\begin{minipage}{.49\textwidth}
		\includegraphics[width=\linewidth]{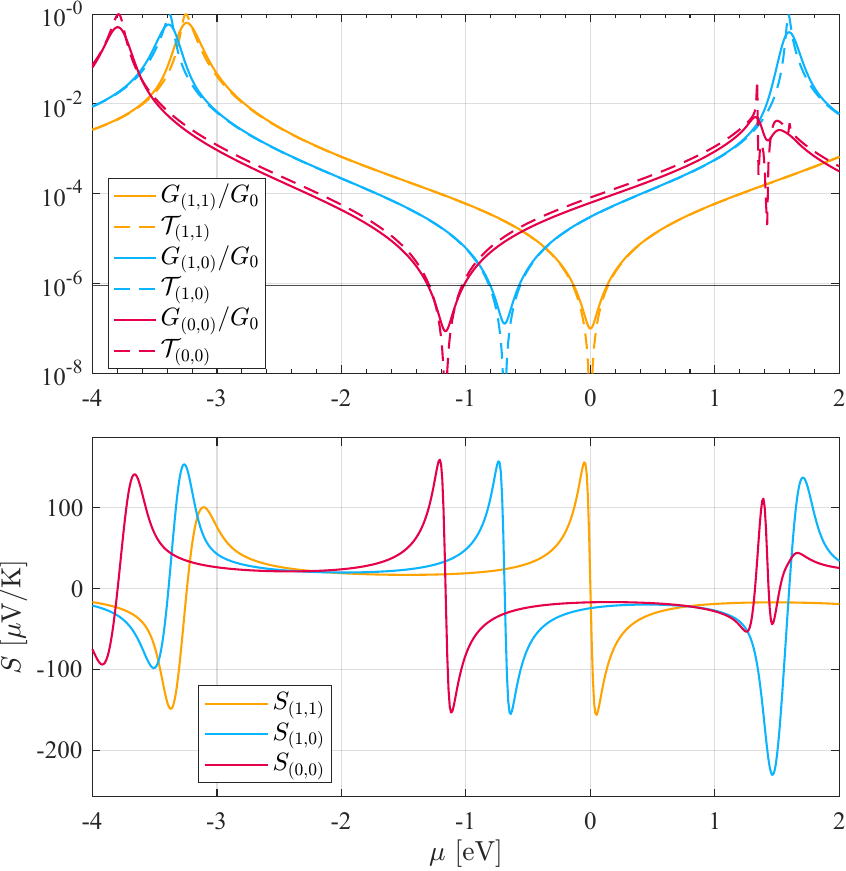}	
		\caption*{{\bf (b)} Many-body MDE theory}
		%		\vspace{.5em}
	\end{minipage}
\caption{Transmission $\mathcal{T}_\pi(E)$, conductance $G(E)$, and thermopower $S(E)$ for \texttt{NAND} junctions computed with {\bf (a)} H\"uckel tight-binding theory and {\bf (b)} many-body MDE theory. The $(0,1)$ and $(1,0)$ inputs are symmetry-equivalent. Energies are referenced to the effective carbon onsite potential, with thiol terminations incorporated via renormalized tunnel couplings. A background $\sigma$-channel (black line) is included using ${\cal T}_\sigma = A_\sigma e^{-\beta_\sigma L}$ with $\beta_\sigma=1$, $A_\sigma=10^{-4}$ (see Appendix), but it does not significantly obscure the interference features in this case. Thermopower exhibits sharp peaks at each transmission node, reaching $\pi/\sqrt{3}\,(k_B/e)\approx 156~\mu$V/K. Both models recover \texttt{NAND} functionality, but correlations reverse the nodal trend: in the H\"uckel limit the $(0,0)$ node shifts upward, whereas in the many-body treatment it shifts downward, underscoring the sensitivity of interference to electronic correlations and symmetry. Conductance is normalized to $G_0=2e^2/h$ and $T_0=300$~K.}

     % \end{adjustwidth}
	\label{fig:NAND_thermo_manybody}
\end{figure}

% C:\SynologyDrive\!Research\2025_0829-Molecular_Circuit\plot_NAND_v2.m
\begin{figure}[tb]
	\centering
    %\isPreprints{}{% This command is only used for ``preprints''.
% \begin{adjustwidth}{-\extralength}{0cm}
	\begin{minipage}{.49\textwidth}
		\includegraphics[width=\linewidth]{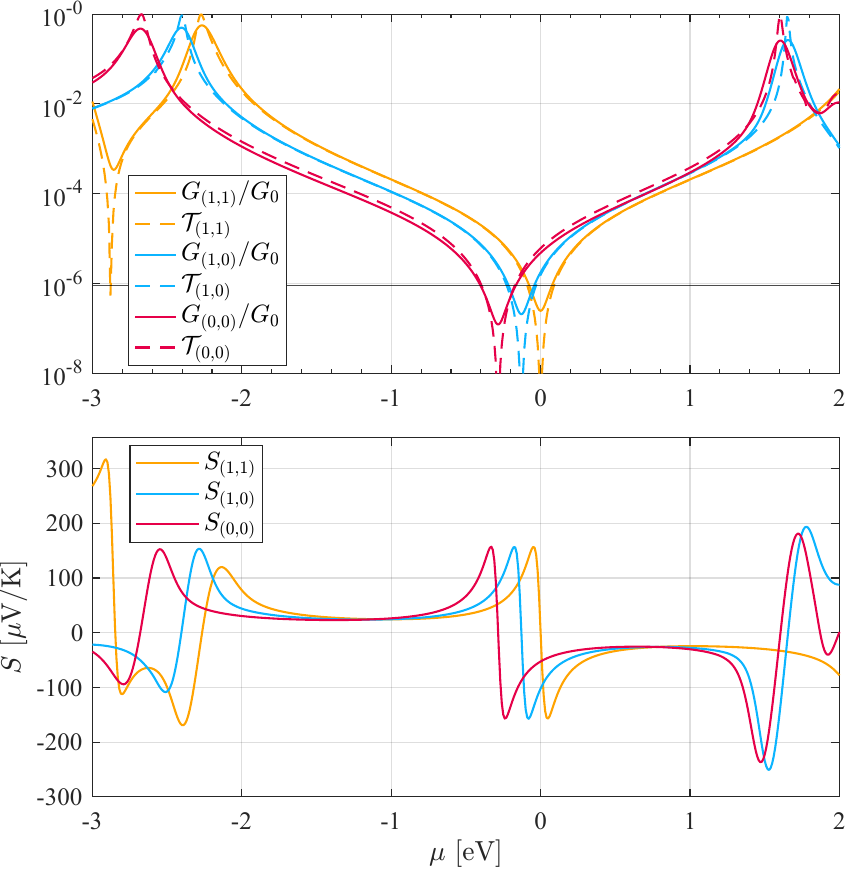}	
%		\caption*{{\bf (a)} Hubbard ($U_{nm} = \delta_{nm} U_0$) many-body theory}
		%		\vspace{.5em}
	\end{minipage}
\caption{Transmission $\mathcal{T}_\pi$, conductance $G$, and thermopower $S$ for \texttt{NAND} junctions computed within the many-body MDE theory including only local Coulomb interactions ($U_{nm} = \delta_{nm}$9.69~eV, i.e.~the Hubbard model).  
As in Fig.~\ref{fig:NAND_thermo_manybody}, the $(0,1)$ and $(1,0)$ inputs yield identical transport owing to molecular symmetry.  
Energies are referenced to the effective carbon onsite potential, and a weak $\sigma$-channel background is shown for comparison.  
Relative to the independent-electron (H\"uckel) case, both the Hubbard and long-range Coulomb descriptions predict that the $(0,0)$ and $(1,0)$ nodes shift downward in energy, underscoring the role of electronic correlation.  
The Hubbard model further restores nearly uniform nodal separations while retaining the overall energetic trends of the full Coulomb calculation, illustrating the distinct influences of local versus nonlocal interactions on logical response.  
Conductance is normalized to $G_0 = 2e^2/h$ and calculations correspond to $T_0=300$~K.}

	\label{fig:NAND_hubbard}
\end{figure}

% C:\SynologyDrive\!Research\2025_0829-Molecular_Circuit\plot_S_discriminant_v2.m
\begin{figure}[tb]
	\centering
    %\isPreprints{}{% This command is only used for ``preprints''.
% \begin{adjustwidth}{-\extralength}{0cm}
	\begin{minipage}{.32\textwidth}
		\includegraphics[width=\linewidth]{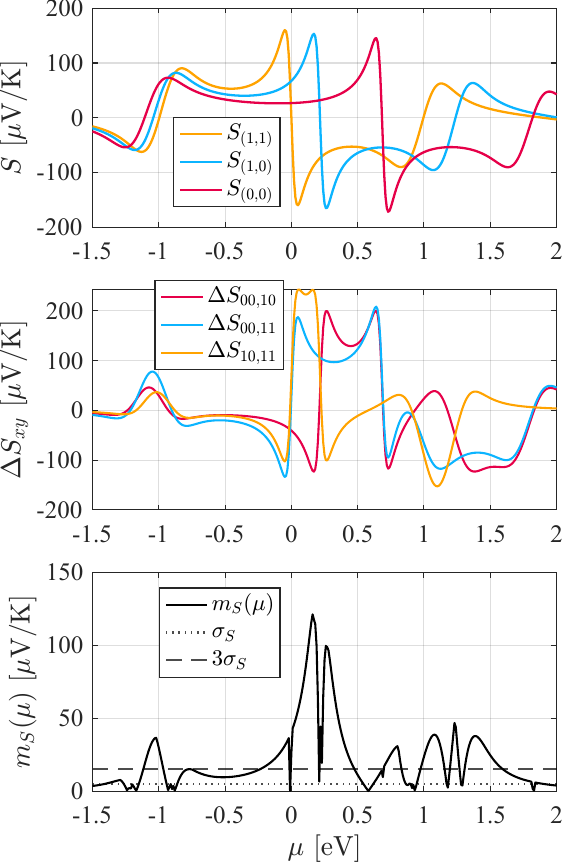}	
		\caption*{{\bf (a)} Single-particle theory}
		%		\vspace{.5em}
	\end{minipage}
	\begin{minipage}{.32\textwidth}
		\includegraphics[width=\linewidth]{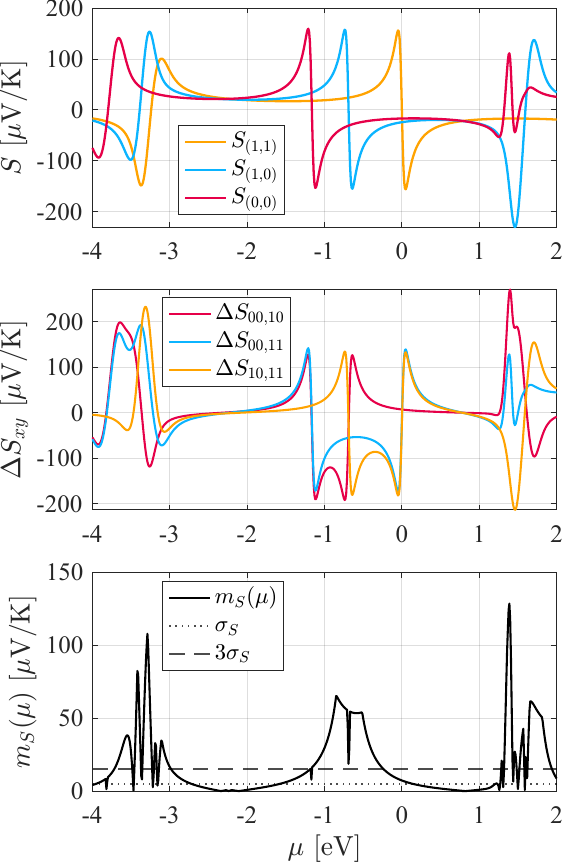}	
		\caption*{{\bf (b)} Many-body theory}
		%		\vspace{.5em}
	\end{minipage}
    \begin{minipage}{.32\textwidth}
		\includegraphics[width=\linewidth]{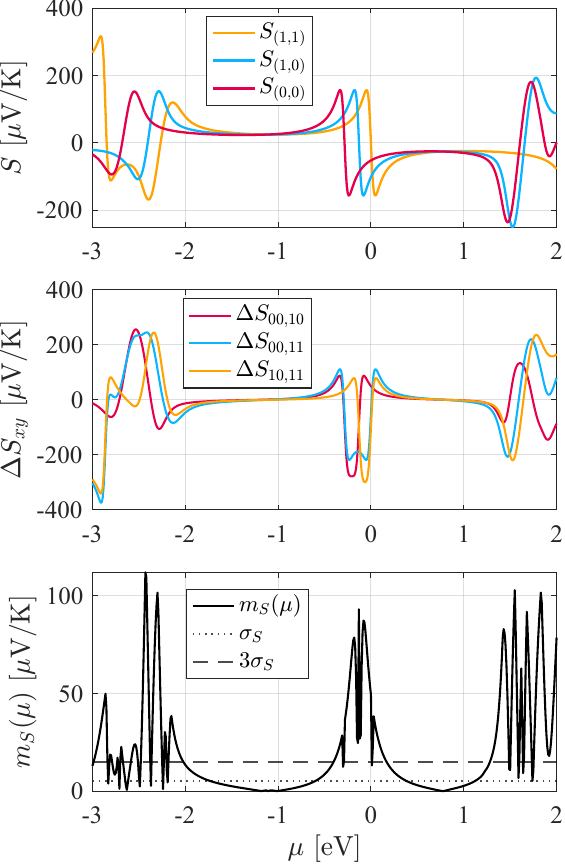}	
		\caption*{{\bf (c)} Many-body (Hubbard model)}
		%		\vspace{.5em}
	\end{minipage}

    \caption{Calculated thermopower $S$, Seebeck contrast $\Delta S$, and discrimination margin $m_S$ for iso-PA-based \texttt{NAND} molecules as a function of chemical potential $\mu$, obtained using three levels of theory:  
{\bf (a)} single-particle H\"uckel tight-binding theory,  
{\bf (b)} many-body MDE theory, and  
{\bf (c)} MDE theory with local Coulomb interactions (Hubbard model with $U_{nm} = U_0 \delta_{nm}$).  
As indicated in the lower panels of each subfigure, the logical output can be discriminated among \texttt{NAND} inputs $(0,0)$, $(1,0)$, and $(1,1)$ by evaluating the thermopower response.  
While all approaches capture interference-induced nodes that underpin logical discrimination, the position, splitting, and magnitude of the thermopower peaks differ substantially, underscoring the sensitivity of readout fidelity to the underlying electronic structure treatment. Horizontal dotted and dashed lines correspond to one and three times a representative experimental uncertainty $\sigma_S$=5$\mu$V/K, respectively, chosen within the $1{-}10~\mu$V/K range established by single-molecule thermopower measurements~\cite{reddy2007thermoelectricity,evangeli2013engineering,rincon2016thermopower}. Transport calculations are for junctions operating at room temperature, $T_0$=300K.}
	\label{fig:NAND_discrim}
\end{figure}

\section{Results}
\label{sec:results}

%\JPB{We begin with}  
\JPB{First,} 
a minimal \texttt{NAND} motif constructed from alkynyl-modified iso-poly(acetylene) (iso-PA) backbones \JPB{is considered}. Logical inputs are encoded chemically: an \ce{NH2} cap denotes bit $0$, detuning the branch from resonance, while a vinyl group (\ce{C=C}) denotes bit $1$, bringing the branch near resonance. The schematic \texttt{NAND} trees and their molecular analogues are shown in Fig.~\ref{fig:NAND_schematic}, where specific substituents implement the input states $(0,0)$, $(0,1)$, and $(1,1)$. In each case, the logical output is read directly from the transmission spectrum near the nodal energy: high transmission corresponds to output $1$, while suppressed transmission signals output $0$. The alkynyl linkers act as insulating standoffs, reducing the $\sigma$-channel transport \JPB{which can obscure interference effects in the $\pi$-system's transport}
%are  notthereby isolating the $\pi$-system interference \JPB{ensuring it dominates the transport} %that governs 
%the low-energy response 
(see Appendix). This convention mirrors the edge-encoding logic of \texttt{NAND} trees, where branch substitution toggles the effective self-energy relative to the reference energy.

To assess this behavior, \JPB{the transport was} computed using many-body MDE theory, which retains the full spectrum of charged and excited states. In the limit $U_{nm}=0$, these calculations reduce to the independent-electron (H\"uckel) tight-binding description. All molecular geometries are fixed at their optimized structures, including thiols and terminal hydrogens for chemical stability. The electrodes are coupled symmetrically with $\Gamma_L$=$\Gamma_R$=0.5eV, and the operating temperature is set to $T_0$=300K. Sulfur anchoring groups are included only through these couplings, allowing direct comparison with prior studies~\cite{jensen2019molecular}. On the vacuum scale, the Au Fermi level lies $\sim 1$~eV below the carbon onsite energy~\cite{rankin2009crc}, but to avoid dependence on a specific alignment %we report 
the full \JPB{spectra are shown}, which can be reinterpreted for alternative linkers or electrode materials.

% The transmission $\mathcal{T}$, conductance $G$, and thermopower $S$ of the $\pi$-system of the iso-PDA-based \texttt{NAND} junctions for the $(0,0)$, $(1,0)$, and $(1,1)$ inputs are using H\"uckel and many-body theories are shown in Fig.~\ref{fig:\texttt{NAND}_thermo_manybody}, respectively. 
The transmission $\mathcal{T}$, conductance $G$, and Seebeck coefficient $S$ for the $\pi$-system of iso-PA \texttt{NAND} junctions are shown in Fig.~\ref{fig:NAND_thermo_manybody}, for input states $(0,0)$, $(1,0)$, and $(1,1)$, as obtained from H\"uckel (Fig.~\ref{fig:NAND_thermo_manybody}a) and many-body MDE calculations (Fig.~\ref{fig:NAND_thermo_manybody}b).
%
% The transmission $\mathcal{T}$, conductance $G$, and thermopower $S$ of the 
% $\pi$-system in iso-PDA \texttt{NAND} junctions for the $(0,0)$, $(1,0)$, and $(1,1)$ 
% inputs are shown in Fig.~\ref{fig:\texttt{NAND}_thermo_manybody}a and b, as obtained from 
% H\"uckel and many-body calculations, respectively.
In the H\"uckel theory results, the expected shifted-node pattern appears: the $(1,1)$ input pins the node near the reference energy (here $\mu=0$), while $(1,0)$ and $(0,0)$ shift the node depending on branch parity and coupling. In linear response with frozen geometries, the $(0,1)$-- and $(1,0)$-junction responses are identical. These results closely match the tight-binding and density functional theory (DFT) calculations of Ref.~\citenum{jensen2019molecular}, which employed longer alkynyl and phenyl linkers, suggesting that at the effective single-particle level the stub network, rather than the precise backbone, controls the interference pattern. As expected, each transmission node generates a conductance dip accompanied by a thermopower enhancement~\cite{bergfield2009thermoelectric,bergfield2010giant}, rendering $S$ an especially sensitive discriminator of logical states.

Importantly, while ``high'' and ``low'' transmission are relative quantities that depend strongly on electrode coupling and linker chemistry, the thermopower provides a more universal readout: because it probes the slope of $\mathcal{T}$ rather than its magnitude, $S$ remains robust to changes in linker length or contact geometry~\cite{bennett2024quantum}. Thus the qualitative features in Fig.~\ref{fig:NAND_thermo_manybody} reproduce earlier tight-binding predictions, despite the shorter linkers employed here, reinforcing the view that the molecular backbone acts largely as a passive conduit, while the stub-system self-energies control the dominant interference physics.

When electron–electron interactions are included, the molecular spectrum is renormalized by the Coulomb self-energy, as illustrated in Fig.~\ref{fig:NAND_thermo_manybody}b. Dynamic correlations increase charging energies, shift and broaden resonances, and modify effective couplings. \JPB{In our iso-PA junctions the Coulomb matrix $U_{nm}$ is long-ranged ($U_{nn}$=9.69~eV onsite, with values $\sim 1.5$~eV even between the most separated sites), so electrons throughout the molecule all interact strongly. These long-range interactions reshape the relative phases of transport amplitudes, thereby altering quantum interference features such as transmission nodes on a global scale.}
% At the same time, the relative phases of transport amplitudes can change substantially, reshaping interference patterns on a global scale. 
% %even distant orbita that distant far apart orbitals nominally far apart become strongly coupled.} 
% These % long-range 
% interactions \JPB{affect the phase of the transport amplitudes and can therefore} %restructures interference beyond the local scale and can, in principle, 
% profoundly alter QI features such as transmission nodes.

Despite these renormalizations, the qualitative \texttt{NAND}-like response of the non-interacting case is retained: the $(1,1)$ junction still exhibits a node at $\mu = 0$, while the other junctions show shifted nodal structure. However, two key differences emerge. First, the nodal shifts invert: in the many-body spectrum the $(0,0)$ and $(1,0)$ nodes move downward in energy rather than upward. This inversion underscores the sensitivity of nodal positions not only to orbital charging and molecular conformation, which are included in the H\"uckel description, but also to the correlated matrix elements that connect many-body states.
Second, chemical substitutions exert a global influence, for example modifying the HOMO–LUMO gap and reshaping resonance line shapes across the spectrum. Because these changes arise from the structure of electron-electron interactions, they cannot, in general, be adjusted arbitrarily. This highlights the intrinsic challenge of engineering a molecular Hamiltonian: the effective parameters are dictated not only by connectivity and local energies, but by the fundamental structure of physical and chemical laws \JPB{which often involve less intuitive non-local and multi-particle properties.}

To better understand the role of non-local interactions, we restrict the Coulomb term to a purely local form, $U_{nm}=U_0\delta_{nm}$ with $U_0=9.69$ eV.  
This reduces the description to a Hubbard model of the molecule, whose transport is shown in Fig.~\ref{fig:NAND_hubbard}.  
In this limit the symmetry of the problem simplifies, with only onsite interactions retained, yet the resulting trends follow the full many-body calculation more closely than the non-interacting H\"uckel theory.  
% The separation between nodes becomes smaller and more uniform, compared to the non-local interaction spectrum, but their positions still shift in response to branch substitution with the same trend as when interactions spanned all sites.  
The separation between nodes becomes smaller and more uniform, though their positions continue to shift under branch substitution as in the full non-local interaction spectrum.
This comparison highlights an important point: the location of interference nodes, and thus the fidelity of logical readout, depends not only on geometrical connectivity but on the deeper structure of the correlated Hilbert space.  
Long-range matrix elements couple orbitals across the entire molecule and reshape interference globally, yet even local (Hubbard-like) interactions alter nodal positions and modify the logical response.  
% Molecular logic, therefore, is not generally a function of tight-binding connectivity alone; instead it requires a more complete description of multi-particle correlations ... ?many-body description including electronic correlations are actively sculpt the effective spectrum and ultimately determine the viability of logical operations.
% Molecular logic, therefore, cannot always be expressed as a function of tight-binding connectivity alone; instead it generally requires a careful treatment of many-body correlations and the Hilbert space that host them.
Molecular logic, therefore, cannot always be expressed solely in terms of tight-binding connectivity; it generally requires a careful treatment of many-body correlations and the Hilbert space that hosts them.

% Molecular logic, therefore, cannot always be reduced to a function of tight-binding connectivity alone, instead our results indicate it generally requires a careful treatment of the multi-particle correlations

% instead it requires a full many-body description in which electronic correlations actively sculpt the effective spectrum and ultimately determine the viability of logical operations.

\subsection{Thermopower discrimination of logic states}

To quantify how well different logical inputs can be distinguished, a Seebeck–contrast metric \JPB{was introduced}. For inputs $x$ and $y$,
\begin{equation}
    \Delta S_{xy}(\mu) = S_x(\mu) - S_y(\mu),
\end{equation}
and the \emph{margin} is defined as
\begin{equation}
    m_S(\mu) = \min_{x \neq y} \big|\Delta S_{xy}(\mu)\big|.
\end{equation}
The margin $m_S$ captures the worst–case separation among all inputs in an experimentally robust observable. It is maximized when the operating energy lies close to, but not exactly at, a transmission node: for a quadratic node the peak thermopower is $(\pi/\sqrt{3})(k_B/e) \approx 156~\mu\mathrm{V/K}$ when $|\mu-\mu_{\rm node}| = (\pi/\sqrt{3})k_B T_0$~\cite{bergfield2009thermoelectric,bennett2024quantum}. Nodes separated by significantly less than this scale are therefore hard to resolve using $S$ alone, although joint statistics of $G$ and $S$ can still enable identification~\cite{bergfield2024identifying}.

The thermopower $S$, the pairwise contrasts $\Delta S_{xy}$, and the margin $m_S$ for H\"uckel, many-body (MDE), and Hubbard-level calculations are shown in Fig.~\ref{fig:NAND_discrim}. Horizontal dotted and dashed lines mark one and three times a representative experimental uncertainty, $\sigma_S = 5~\mu\mathrm{V/K}$, chosen conservatively from the $1$–$10~\mu\mathrm{V/K}$ range reported in single-molecule thermopower studies~\cite{reddy2007thermoelectricity,evangeli2013engineering,rincon2016thermopower}. We adopt $m_S > 3\sigma_S$ as a criterion for reliable discrimination: in this regime even the closest pair of states remains separable, and the corresponding worst-case error probability
\begin{equation}
    P_{\rm err}^{\rm (worst)}(\mu) \approx 
    \frac{1}{2}\,\mathrm{erfc}\!\left(\frac{m_S(\mu)}{2\sqrt{2}\,\sigma_S}\right)
\end{equation}
falls below $\sim\!0.3\%$. By contrast, regions with $m_S \lesssim \sigma_S$ are noise-limited and should be avoided. As seen in the $m_S$ panels of Fig.~\ref{fig:NAND_discrim}, each logical state can be reliably discriminated over wide ranges of chemical potential, consistent with variations in electrode material or linker alignment.

Assessing the efficacy of molecular \texttt{NAND} gates is necessarily nuanced. Our many-body calculations %often
agree \emph{qualitatively} with the noninteracting H\"uckel and DFT results~\cite{jensen2019molecular}; nodes shift when substituents are modified, showing that simplified models can capture the essentials in some regimes. Quantitatively, however, the nodal energies and even the \emph{direction} of substitution-induced shifts can differ, reflecting the fact that weak, long-range Coulomb interactions couple distant orbitals and affect the phase of transport amplitudes across the molecule. What may look like a local, stub-induced node shift can instead originate from nonlocal correlations among many-body states~\cite{bergfield2011novel}, cautioning against \JPB{over-reliance} on single-particle pictures. In this context, the thermopower provides a robust, chemistry-agnostic readout: by probing the energy derivative of the transmission, $S$ accentuates nodal structure even when conductance signatures blur, serving both as a sensitive diagnostic of correlation-driven rephasing and as a practical route to reliable logic readout in molecular devices.

\section{Conclusions}
\label{sec:conclusions}

We have investigated chemically encoded \texttt{NAND} gates based on modified iso-PA motifs using independent-electron (H\"uckel), Hubbard-level, and fully many-body (MDE) theories of transport. Across all descriptions, interference-induced transmission nodes persist, confirming that \texttt{NAND}-like functionality can, in principle, be realized in single-molecule junctions. At the same time, correlations reshape nodal positions and can even invert substitution trends, exposing the limits of simplified one-body %“stub-resonator’’ 
models.
%and turning these junctions into natural diagnostics of correlation strength.

While node identification lies at the heart of scattering-based readout, its experimental realization is nontrivial: finite signal-to-noise obscures shallow dips and contact chemistry can induce orders-of-magnitude conductance variations. By contrast, the thermopower $S$ provides a chemically robust discriminator of logical states. Because $S$ probes the energy derivative of the transmission, it preserves contrast when conductance signatures blur. We quantified this with the Seebeck margin $m_S$, and by linking $m_S$ to a representative experimental uncertainty $\sigma_S$ we obtained a practical criterion for reliable readout ($m_S>3\sigma_S$, implying a worst-case error probability below $\sim 0.3\%$). This connects logical fidelity directly to measurable transport observables.

% Viewed in this light, molecule-based quantum \texttt{NAND} devices embody both the promise and the difficulty of molecular information processing: they show that logical operations can, in principle, be inscribed directly into chemical structure, while reminding us that Hamiltonian engineering is ultimately constrained by the laws of physics and chemistry.  
% Robust function demands a delicate balance of quantum interference with thermodynamic constraints, and the challenge ahead is to transform this balance from a subtlety of theory into a foundation for scalable molecular logic. 

This work illustrates that molecular junctions are more than simple physical realizations of abstract logic gates: they are laboratories to investigate the fundamental interplay between quantum coherence, correlation, and computation. Seen from this perspective, interactions become a design parameter to work within, capable of stabilizing, destabilizing, or even \emph{generating} nodes with no purely geometric origin, thereby suggesting novel routes to realize logic gates. The remaining challenge is to translate these insights into chemically scalable motifs and gate-tunable junctions where interaction-shaped interference is engineered, not endured.

\funding{This research was graciously supported by the National Science Foundation under award number QIS-2412920.}

%\section*{Data Availability Statement}
%The data that support the findings of
%this study are available from the
%corresponding author upon reasonable
%request.
%

\institutionalreview{Not applicable}

\informedconsent{Not applicable}

\dataavailability{The original contributions presented in the study are included in the article, further inquiries can be directed to the corresponding author.}

\conflictsofinterest{The author declares no conflicts of interest.}

% %%%%%%%%%%%%%%%%%%%%%%%%%%%%%%%%%%%%%%%%%%
% %% Optional
% \appendixtitles{yes} % Leave argument "no" if all appendix headings stay EMPTY (then no dot is printed after "Appendix A"). If the appendix sections contain a heading then change the argument to "yes".
 \appendixstart
 \appendix

\section{The Influence of  \texorpdfstring{$\sigma$}{s}-system Transport}
\label{app:sigma}

The total transmission through a single-molecule junction can be written as the sum of $\pi$- and $\sigma$-channel contributions,
\begin{equation}
{\cal T}_{\rm tot}(E) = {\cal T}_\pi(E) + {\cal T}_\sigma(E).
\end{equation}
For conjugated backbones such as the alkynyl-extended iso-PA structures considered here, ${\cal T}_\sigma(E)$ is nearly energy-independent in the mid-gap window,\cite{solomon2008understanding} and thus contributes a constant background. Within linear response the corresponding Onsager coefficients are
\begin{equation}
\Lfn{0}_\sigma = \frac{\mathcal T_\sigma}{h}, \qquad
\Lfn{1}_\sigma = 0,
\end{equation}
so that the conductance acquires an additive offset while the numerator of the Seebeck coefficient is unchanged. As a result,
\begin{equation}
S_{\rm tot} = S_\pi \,\frac{\Lfn{0}_\pi}{\Lfn{0}_\pi + \Lfn{0}_\sigma},
\label{eq:Stot_sigma}
\end{equation}
showing that $\sigma$ transport invariably suppresses the thermopower by diluting the node-driven $\pi$ response.

Although the $\sigma$-system is formally included via the renormalized parameters of our effective $\pi$-Hamiltonian, its value can be estimated independently from benchmarks on saturated alkanes. Off-resonant $\sigma$ tunneling follows an exponential law~\cite{van2022benchmark,guo2011measurement},
\begin{equation}
  \mathcal{T}_\sigma(L) = A_\sigma e^{-\beta_\sigma L},
\end{equation}
with decay constants $\beta_\sigma \simeq 1.0$–$1.1$ per \ce{CH2} ($\sim 0.8$–$1.0$~\AA$^{-1}$) for thiols and $\beta_\sigma \simeq 0.8$ per \ce{CH2} ($\sim 0.6$~\AA$^{-1}$) for amines. These values reproduce the canonical C3/C6/C8 conductance peaks and are widely used as benchmarks~\cite{li2006conductance,li2008charge,haiss2009impact,venkataraman2006single}.

If one adopts the Au–S prefactor $A_\sigma \approx 0.1$, $\sigma$-channel transmission would already be of order $10^{-4}$ at spans $L \sim 6$–8~\AA, comparable to measured conductances of hexanedithiol and octanedithiol. Such contributions could overwhelm nodal variations in $\mathcal{T}_\pi(E)$ and wash out the associated thermopower enhancements. In practice, however, the effective $A_\sigma$ depends strongly on contact chemistry and geometry. Anchoring groups such as amines, pyridyls, or sp-hybridized alkynyl linkers, as well as top-site or tilted binding configurations, can reduce $A_\sigma$ by one to two orders of magnitude. For the iso-PA-based junctions considered here, the extended sp-rich scaffolds intrinsically suppress $\sigma$ overlap, making them especially attractive for logic applications. In simulations we therefore adopt representative parameters $\beta_\sigma = 1$~\AA$^{-1}$ and $A_\sigma = 10^{-4}$, which yield $\mathcal{T}_\sigma$ values small enough to preserve the diagnostic node-enhanced thermopower.

% \section{Optimization}

% % Point?
% \begin{figure}[tb]
% 	\centering
% 	\includegraphics[width=.3\linewidth]{Figures_final/\texttt{NAND}00.prb}
% 	\includegraphics[width=.3\linewidth]{Figures_final/\texttt{NAND}10.prb}
% 	\includegraphics[width=.3\linewidth]{Figures_final/\texttt{NAND}11.prb}
% \caption{Optimized geometreis}
% 	\label{fig:\texttt{NAND}_geometry}
% \end{figure}

\bibliography{refs_clean}

\begin{thebibliography}{999}

\bibitem[Deutsch and Jozsa(1992)]{deutsch1992rapid}
Deutsch, D.; Jozsa, R.
\newblock Rapid solution of problems by quantum computation.
\newblock {\em Proc. R. Soc. Lond. A.} {\bf 1992}, {\em 439},~553--558.

\bibitem[Shor(1999)]{shor1999polynomial}
Shor, P.W.
\newblock Polynomial-time algorithms for prime factorization and discrete
  logarithms on a quantum computer.
\newblock {\em SIAM review} {\bf 1999}, {\em 41},~303--332.

\bibitem[Childs(2009)]{childs2009universal}
Childs, A.M.
\newblock Universal Computation by Quantum Walk.
\newblock {\em Phys. Rev. Lett.} {\bf 2009}, {\em 102},~180501.
\newblock {\url{https://doi.org/10.1103/PhysRevLett.102.180501}}.

\bibitem[Childs et~al.(2013)Childs, Gosset, and Webb]{childs2013universal}
Childs, A.M.; Gosset, D.; Webb, Z.
\newblock Universal computation by multiparticle quantum walk.
\newblock {\em Science} {\bf 2013}, {\em 339},~791--794.

\bibitem[Venegas-Andraca(2012)]{venegas2012quantum}
Venegas-Andraca, S.E.
\newblock Quantum walks: a comprehensive review.
\newblock {\em Quantum Inf. Process.} {\bf 2012}, {\em 11},~1015--1106.

\bibitem[Childs and Strouse(2011)]{childs2011levinson}
Childs, A.M.; Strouse, D.
\newblock Levinson's theorem for graphs.
\newblock {\em J. Math. Phys.} {\bf 2011}, {\em 52}.

\bibitem[Childs and Gosset(2012)]{childs2012levinson}
Childs, A.M.; Gosset, D.
\newblock Levinson's theorem for graphs II.
\newblock {\em J. Math. Phys.} {\bf 2012}, {\em 53}.

\bibitem[Whitfield et~al.(2010)Whitfield, Rodr{\'\i}guez-Rosario, and
  Aspuru-Guzik]{whitfield2010quantum}
Whitfield, J.D.; Rodr{\'\i}guez-Rosario, C.A.; Aspuru-Guzik, A.
\newblock Quantum stochastic walks: A generalization of classical random walks
  and quantum walks.
\newblock {\em Phys. Rev. A} {\bf 2010}, {\em 81},~022323.

\bibitem[Farhi and Gutmann(1998)]{farhi1998quantum}
Farhi, E.; Gutmann, S.
\newblock Quantum computation and decision trees.
\newblock {\em Phys. Rev. A} {\bf 1998}, {\em 58},~915.

\bibitem[Farhi et~al.(2007)Farhi, Goldstone, and
  Gutmann]{farhi2007quantumalgorithmhamiltoniannand}
Farhi, E.; Goldstone, J.; Gutmann, S.
\newblock A Quantum Algorithm for the Hamiltonian NAND Tree,  2007,
  \href{http://arxiv.org/abs/quant-ph/0702144}{{\normalfont
  [arXiv:quant-ph/quant-ph/0702144]}}.

\bibitem[Hoyer and Spalek(2005)]{hoyer2005lower}
Hoyer, P.; Spalek, R.
\newblock Lower Bounds on Quantum Query Complexity,  2005,
  \href{http://arxiv.org/abs/quant-ph/0509153}{{\normalfont
  [arXiv:quant-ph/quant-ph/0509153]}}.

\bibitem[Saks and Wigderson(1986)]{saks1986probabilistic}
Saks, M.; Wigderson, A.
\newblock Probabilistic Boolean decision trees and the complexity of evaluating
  game trees.
\newblock In Proceedings of the 27th Annual Symposium on Foundations of
  Computer Science (sfcs 1986). IEEE,  1986, pp. 29--38.

\bibitem[Tsuji et~al.(2018)Tsuji, Estrada, Movassagh, and
  Hoffmann]{tsuji2018quantum}
Tsuji, Y.; Estrada, E.; Movassagh, R.; Hoffmann, R.
\newblock Quantum interference, graphs, walks, and polynomials.
\newblock {\em Chem. Rev.} {\bf 2018}, {\em 118},~4887--4911.

\bibitem[Namarvar et~al.(2019)Namarvar, Giraud, Georgeot, and
  Joachim]{namarvar2019quantum}
Namarvar, O.F.; Giraud, O.; Georgeot, B.; Joachim, C.
\newblock Quantum Hamiltonian Computing protocols for molecular electronics
  Boolean logic gates.
\newblock {\em Quantum Sci. Technol.} {\bf 2019}, {\em 4},~035009.

\bibitem[Baytekin and Akkaya(2000)]{baytekin2000molecular}
Baytekin, H.T.; Akkaya, E.U.
\newblock A molecular NAND gate based on Watson- Crick base pairing.
\newblock {\em Organic Letters} {\bf 2000}, {\em 2},~1725--1727.

\bibitem[Erbas-Cakmak et~al.(2018)Erbas-Cakmak, Kolemen, Sedgwick,
  Gunnlaugsson, James, Yoon, and Akkaya]{erbas2018molecular}
Erbas-Cakmak, S.; Kolemen, S.; Sedgwick, A.C.; Gunnlaugsson, T.; James, T.D.;
  Yoon, J.; Akkaya, E.U.
\newblock Molecular logic gates: the past, present and future.
\newblock {\em Chem. Soc. Rev.} {\bf 2018}, {\em 47},~2228--2248.

\bibitem[Kompa and Levine(2001)]{kompa2001molecular}
Kompa, K.; Levine, R.
\newblock A molecular logic gate.
\newblock {\em P. Natl. Acad. Sci. USA} {\bf 2001}, {\em 98},~410--414.

\bibitem[Evers et~al.(2020)Evers, Koryt{\'a}r, Tewari, and van
  Ruitenbeek]{evers2020advances}
Evers, F.; Koryt{\'a}r, R.; Tewari, S.; van Ruitenbeek, J.M.
\newblock Advances and challenges in single-molecule electron transport.
\newblock {\em Rev. Mod. Phys.} {\bf 2020}, {\em 92},~035001.

\bibitem[Jensen et~al.(2019)Jensen, Jin, Dallaire-Demers, Aspuru-Guzik, and
  Solomon]{jensen2019molecular}
Jensen, P.W.; Jin, C.; Dallaire-Demers, P.L.; Aspuru-Guzik, A.; Solomon, G.C.
\newblock Molecular realization of a quantum NAND tree.
\newblock {\em Quantum Sci. Technol.} {\bf 2019}, {\em 4},~015013.

\bibitem[Soe et~al.(2011)Soe, Manzano, Renaud, de~Mendoza, De~Sarkar, Ample,
  Hliwa, Echavarren, Chandrasekhar, and Joachim]{soe2011manipulating}
Soe, W.H.; Manzano, C.; Renaud, N.; de~Mendoza, P.; De~Sarkar, A.; Ample, F.;
  Hliwa, M.; Echavarren, A.M.; Chandrasekhar, N.; Joachim, C.
\newblock Manipulating molecular quantum states with classical metal atom
  inputs: demonstration of a single molecule NOR logic gate.
\newblock {\em ACS Nano} {\bf 2011}, {\em 5},~1436--1440.

\bibitem[Joachim et~al.(2012)Joachim, Renaud, and Hliwa]{joachim2012different}
Joachim, C.; Renaud, N.; Hliwa, M.
\newblock The different designs of molecule logic gates,  2012.

\bibitem[Renaud and Joachim(2008)]{renaud2008design}
Renaud, N.; Joachim, C.
\newblock Design and stability of NOR and NAND logic gates constructed with
  three quantum states.
\newblock {\em Phys. Rev. A} {\bf 2008}, {\em 78},~062316.

\bibitem[Joachim et~al.(2000)Joachim, Gimzewski, and
  Aviram]{joachim2000electronics}
Joachim, C.; Gimzewski, J.; Aviram, A.
\newblock Electronics using hybrid-molecular and mono-molecular devices.
\newblock {\em Nature} {\bf 2000}, {\em 408},~541--548.

\bibitem[Dridi et~al.(2015)Dridi, Julien, Hliwa, and
  Joachim]{dridi2015mathematics}
Dridi, G.; Julien, R.; Hliwa, M.; Joachim, C.
\newblock The mathematics of a quantum Hamiltonian computing half adder Boolean
  logic gate.
\newblock {\em Nanotechnology} {\bf 2015}, {\em 26},~344003.

\bibitem[Dridi et~al.(2018)Dridi, Namarvar, and Joachim]{dridi2018qubits}
Dridi, G.; Namarvar, O.F.; Joachim, C.
\newblock Qubits and quantum Hamiltonian computing performances for operating a
  digital Boolean 1/2-adder.
\newblock {\em Quantum Sci. Technol.} {\bf 2018}, {\em 3},~025005.

\bibitem[Joachim et~al.(2005)Joachim, Duchemin, Fiur{\'a}{\v{s}}ek, and
  Cerf]{joachim2005hamiltonian}
Joachim, C.; Duchemin, I.; Fiur{\'a}{\v{s}}ek, J.; Cerf, N.
\newblock Hamiltonian logic gates: computing inside a molecule.
\newblock {\em Int. J. Nanosci.} {\bf 2005}, {\em 4},~107--118.

\bibitem[Bergfield et~al.(2011)Bergfield, Solomon, Stafford, and
  Ratner]{bergfield2011novel}
Bergfield, J.P.; Solomon, G.C.; Stafford, C.A.; Ratner, M.A.
\newblock Novel quantum interference effects in transport through molecular
  radicals.
\newblock {\em Nano Lett.} {\bf 2011}, {\em 11},~2759--2764.

\bibitem[Barr and Stafford(2013)]{barr2013transmission}
Barr, J.D.; Stafford, C.A.
\newblock On Transmission Node Structure in Interacting Systems,  2013,
  \href{http://arxiv.org/abs/1303.3618}{{\normalfont
  [arXiv:cond-mat.mes-hall/1303.3618]}}.

\bibitem[Pedersen et~al.(2014)Pedersen, Strange, Leijnse, Hedeg{\aa}rd,
  Solomon, and Paaske]{pedersen2014quantum}
Pedersen, K.G.; Strange, M.; Leijnse, M.; Hedeg{\aa}rd, P.; Solomon, G.C.;
  Paaske, J.
\newblock Quantum interference in off-resonant transport through single
  molecules.
\newblock {\em Phys. Rev. B} {\bf 2014}, {\em 90},~125413.

\bibitem[Arroyo et~al.(2013)Arroyo, Tarkuc, Frisenda, Seldenthuis, Woerde,
  Eelkema, Grozema, and van~der Zant]{arroyo2013signatures}
Arroyo, C.R.; Tarkuc, S.; Frisenda, R.; Seldenthuis, J.S.; Woerde, C.H.;
  Eelkema, R.; Grozema, F.C.; van~der Zant, H.S.
\newblock Signatures of Quantum Interference Effects on Charge Transport
  Through a Single Benzene Ring.
\newblock {\em Angew. Chem.} {\bf 2013}, {\em 125},~3234--3237.

\bibitem[Gu{\'e}don et~al.(2012)Gu{\'e}don, Valkenier, Markussen, Thygesen,
  Hummelen, and van~der Molen]{guedon2012observation}
Gu{\'e}don, C.M.; Valkenier, H.; Markussen, T.; Thygesen, K.S.; Hummelen, J.C.;
  van~der Molen, S.J.
\newblock Observation of quantum interference in molecular charge transport.
\newblock {\em Nat. Nanotechnol.} {\bf 2012}, {\em 7},~305--309.

\bibitem[Li et~al.(2019)Li, Buerkle, Li, Rostamian, Wang, Wang, Bowler,
  Miyazaki, Xiang, Asai, et~al.]{li2019gate}
Li, Y.; Buerkle, M.; Li, G.; Rostamian, A.; Wang, H.; Wang, Z.; Bowler, D.R.;
  Miyazaki, T.; Xiang, L.; Asai, Y.;  et~al.
\newblock Gate controlling of quantum interference and direct observation of
  anti-resonances in single molecule charge transport.
\newblock {\em Nat. Mater.} {\bf 2019}, {\em 18},~357--363.

\bibitem[Liu et~al.(2018)Liu, Huang, Wang, and Hong]{liu2018quantum}
Liu, J.; Huang, X.; Wang, F.; Hong, W.
\newblock Quantum interference effects in charge transport through
  single-molecule junctions: detection, manipulation, and application.
\newblock {\em Acc. Chem. Res.} {\bf 2018}, {\em 52},~151--160.

\bibitem[Nairz et~al.(2003)Nairz, Arndt, and Zeilinger]{nairz2003quantum}
Nairz, O.; Arndt, M.; Zeilinger, A.
\newblock Quantum interference experiments with large molecules.
\newblock {\em Am. J. Phys.} {\bf 2003}, {\em 71},~319--325.

\bibitem[Aradhya et~al.(2012)Aradhya, Meisner, Krikorian, Ahn, Parameswaran,
  Steigerwald, Nuckolls, and Venkataraman]{aradhya2012dissecting}
Aradhya, S.V.; Meisner, J.S.; Krikorian, M.; Ahn, S.; Parameswaran, R.;
  Steigerwald, M.L.; Nuckolls, C.; Venkataraman, L.
\newblock Dissecting contact mechanics from quantum interference in
  single-molecule junctions of stilbene derivatives.
\newblock {\em Nano Lett.} {\bf 2012}, {\em 12},~1643--1647.

\bibitem[Wang et~al.(2020)Wang, Bennett, Ismael, Wilkinson, Hamill, White,
  Grace, Kolosov, Albrecht, Robinson, et~al.]{wang2020scale}
Wang, X.; Bennett, T.L.; Ismael, A.; Wilkinson, L.A.; Hamill, J.; White, A.J.;
  Grace, I.M.; Kolosov, O.V.; Albrecht, T.; Robinson, B.J.;  et~al.
\newblock Scale-up of room-temperature constructive quantum interference from
  single molecules to self-assembled molecular-electronic films.
\newblock {\em J. Am. Chem. Soc.} {\bf 2020}, {\em 142},~8555--8560.

\bibitem[Liu et~al.(2023)Liu, Ismael, Al-Jobory, and
  Lambert]{liu2023signatures}
Liu, S.X.; Ismael, A.K.; Al-Jobory, A.; Lambert, C.J.
\newblock Signatures of room-temperature quantum interference in molecular
  junctions.
\newblock {\em Acc. Chem. Res.} {\bf 2023}, {\em 56},~322--331.

\bibitem[Bergfield and Stafford(2009)]{bergfield2009many}
Bergfield, J.P.; Stafford, C.A.
\newblock Many-body theory of electronic transport in single-molecule
  heterojunctions.
\newblock {\em Phys. Rev. B} {\bf 2009}, {\em 79},~245125.
\newblock {\url{https://doi.org/10.1103/PhysRevB.79.245125}}.

\bibitem[Bergfield et~al.(2010)Bergfield, Solis, and
  Stafford]{bergfield2010giant}
Bergfield, J.P.; Solis, M.A.; Stafford, C.A.
\newblock Giant thermoelectric effect from transmission supernodes.
\newblock {\em ACS Nano} {\bf 2010}, {\em 4},~5314--5320.

\bibitem[Bennett et~al.(2024)Bennett, Hendrickson, and
  Bergfield]{bennett2024quantum}
Bennett, R.X.; Hendrickson, J.R.; Bergfield, J.P.
\newblock Quantum Interference Enhancement of the Spin-Dependent Thermoelectric
  Response.
\newblock {\em ACS Nano} {\bf 2024}, {\em 18},~11876--11885.

\bibitem[Erdogan and
  Bergfield(2025)]{erdogan2025dephasingfailsthermodynamicconsequences}
Erdogan, E.; Bergfield, J.P.
\newblock When Dephasing Fails: Thermodynamic Consequences of Decoherence
  Models in Quantum Transport,  2025,
  \href{http://arxiv.org/abs/2508.20343}{{\normalfont
  [arXiv:cond-mat.mes-hall/2508.20343]}}.

\bibitem[Miao et~al.(2018)Miao, Xu, Skripnik, Cui, Wang, Pedersen, Leijnse,
  Pauly, Waarnmark, Meyhofer, et~al.]{miao2018influence}
Miao, R.; Xu, H.; Skripnik, M.; Cui, L.; Wang, K.; Pedersen, K.G.; Leijnse, M.;
  Pauly, F.; Waarnmark, K.; Meyhofer, E.;  et~al.
\newblock Influence of quantum interference on the thermoelectric properties of
  molecular junctions.
\newblock {\em Nano Lett.} {\bf 2018}, {\em 18},~5666--5672.

\bibitem[Bergfield(2024)]{bergfield2024identifying}
Bergfield, J.P.
\newblock Identifying Quantum Interference Effects from Joint
  Conductance--Thermopower Statistics.
\newblock {\em Nano Letters} {\bf 2024}, {\em 24},~15110--15117.

\bibitem[Hamill et~al.(2023)Hamill, Ismael, Al-Jobory, Bennett, Alshahrani,
  Wang, Akers-Douglas, Wilkinson, Robinson, Long, et~al.]{hamill2023quantum}
Hamill, J.M.; Ismael, A.; Al-Jobory, A.; Bennett, T.L.; Alshahrani, M.; Wang,
  X.; Akers-Douglas, M.; Wilkinson, L.A.; Robinson, B.J.; Long, N.J.;  et~al.
\newblock Quantum interference and contact effects in the thermoelectric
  performance of anthracene-based molecules.
\newblock {\em J. Phys. Chem. C} {\bf 2023}, {\em 127},~7484--7491.

\bibitem[{Finch} et~al.(2009){Finch}, {Garc{\'{\i}}a-Su{\'a}rez}, and
  {Lambert}]{Finch09}
{Finch}, C.M.; {Garc{\'{\i}}a-Su{\'a}rez}, V.M.; {Lambert}, C.J.
\newblock {Giant thermopower and figure of merit in single-molecule devices}.
\newblock {\em Phys. Rev. B} {\bf 2009}, {\em 79},~033405.

\bibitem[Ke et~al.(2008)Ke, Yang, and Baranger]{ke2008quantum}
Ke, S.H.; Yang, W.; Baranger, H.U.
\newblock Quantum-interference-controlled molecular electronics.
\newblock {\em Nano Lett.} {\bf 2008}, {\em 8},~3257--3261.

\bibitem[Cardamone et~al.(2006)Cardamone, Stafford, and
  Mazumdar]{cardamone2006controlling}
Cardamone, D.M.; Stafford, C.A.; Mazumdar, S.
\newblock Controlling quantum transport through a single molecule.
\newblock {\em Nano Lett.} {\bf 2006}, {\em 6},~2422--2426.

\bibitem[Hansen et~al.(2009)Hansen, Solomon, Andrews, and
  Ratner]{hansen2009interfering}
Hansen, T.; Solomon, G.C.; Andrews, D.Q.; Ratner, M.A.
\newblock Interfering pathways in benzene: An analytical treatment.
\newblock {\em J. Chem. Phys.} {\bf 2009}, {\em 131}.

\bibitem[Markussen et~al.(2010)Markussen, Stadler, and
  Thygesen]{markussen2010relation}
Markussen, T.; Stadler, R.; Thygesen, K.S.
\newblock The relation between structure and quantum interference in single
  molecule junctions.
\newblock {\em Nano Lett.} {\bf 2010}, {\em 10},~4260--4265.

\bibitem[Solomon et~al.(2008)Solomon, Andrews, Goldsmith, Hansen, Wasielewski,
  Van~Duyne, and Ratner]{Solomon08}
Solomon, G.C.; Andrews, D.Q.; Goldsmith, R.H.; Hansen, T.; Wasielewski, M.R.;
  Van~Duyne, R.P.; Ratner, M.A.
\newblock Quantum Interference in Acyclic Systems: Conductance of
  Cross-Conjugated Molecules.
\newblock {\em J. Am. Chem. Soc.} {\bf 2008}, {\em 130},~17301--17308.

\bibitem[Solomon et~al.(2009)Solomon, Andrews, Van~Duyne, and
  Ratner]{solomon2009electron}
Solomon, G.C.; Andrews, D.Q.; Van~Duyne, R.P.; Ratner, M.A.
\newblock Electron transport through conjugated molecules: When the $\pi$
  system only tells part of the story.
\newblock {\em ChemPhysChem} {\bf 2009}, {\em 10},~257--264.

\bibitem[Lunde and Flensberg(2005)]{lunde2005mott}
Lunde, A.M.; Flensberg, K.
\newblock On the Mott formula for the thermopower of non-interacting electrons
  in quantum point contacts.
\newblock {\em J. Phys. Condens. Matter} {\bf 2005}, {\em 17},~3879.

\bibitem[Bergfield and Stafford(2009)]{bergfield2009thermoelectric}
Bergfield, J.P.; Stafford, C.A.
\newblock Thermoelectric Signatures of Coherent Transport in Single-Molecule
  Heterojunctions.
\newblock {\em Nano Lett.} {\bf 2009}, {\em 9},~3072--3076,
  \href{http://arxiv.org/abs/https://doi.org/10.1021/nl901554s}{{\normalfont
  [https://doi.org/10.1021/nl901554s]}}.
\newblock PMID: 19610653, {\url{https://doi.org/10.1021/nl901554s}}.

\bibitem[Bergfield(2025)]{bergfield2025quantuminterferencesupernodesthermoelectric}
Bergfield, J.P.
\newblock Quantum Interference Supernodes, Thermoelectric Enhancement, and the
  Role of Dephasing.
\newblock {\em Entropy} {\bf 2025}.

\bibitem[Inui et~al.(2018)Inui, Stafford, and Bergfield]{inui2018emergence}
Inui, S.; Stafford, C.A.; Bergfield, J.P.
\newblock Emergence of Fourier's law of heat transport in quantum electron
  systems.
\newblock {\em ACS Nano} {\bf 2018}, {\em 12},~4304--4311.

\bibitem[Solomon et~al.(2011)Solomon, Bergfield, Stafford, and
  Ratner]{solomon2011small}
Solomon, G.C.; Bergfield, J.P.; Stafford, C.A.; Ratner, M.A.
\newblock When ``small'' terms matter: Coupled interference features in the
  transport properties of cross-conjugated molecules.
\newblock {\em Beilstein J. Nanotechnol.} {\bf 2011}, {\em 2},~862--871.

\bibitem[Meir and Wingreen(1992)]{meir1992landauer}
Meir, Y.; Wingreen, N.S.
\newblock Landauer formula for the current through an interacting electron
  region.
\newblock {\em Phys. Rev. Lett.} {\bf 1992}, {\em 68},~2512.

\bibitem[Imry and Landauer(1999)]{imry1999conductance}
Imry, Y.; Landauer, R.
\newblock Conductance viewed as transmission. In {\em More Things in Heaven and
  Earth}; Springer,  1999; pp. 515--525.

\bibitem[Haug and Jauho(1996)]{HaugAndJauhoBook}
Haug, H.; Jauho, A.P.
\newblock {\em Quantum Kinetics in Transport and Optics of Semiconductors};
  Vol. 123, {\em Solid-State Sciences}, Springer,  1996.

\bibitem[Stefanucci and Van~Leeuwen(2013)]{stefanucci2013nonequilibrium}
Stefanucci, G.; Van~Leeuwen, R.
\newblock {\em Nonequilibrium many-body theory of quantum systems: a modern
  introduction}; Cambridge University Press,  2013.

\bibitem[Cuniberti et~al.(2005)Cuniberti, Fagas, and
  Richter]{cuniberti2005introducing}
Cuniberti, G.; Fagas, G.; Richter, K.
\newblock {\em Introducing molecular electronics: A brief overview}; Springer,
  2005.

\bibitem[Brisker et~al.(2008)Brisker, Cherkes, Gnodtke, Jarukanont, Klaiman,
  Koch, Weissman, Volkovich, Toroker, and Peskin]{brisker2008controlled}
Brisker, D.; Cherkes, I.; Gnodtke, C.; Jarukanont, D.; Klaiman, S.; Koch, W.;
  Weissman, S.; Volkovich, R.; Toroker, M.C.; Peskin, U.
\newblock Controlled electronic transport through branched molecular
  conductors.
\newblock {\em Mol. Phys.} {\bf 2008}, {\em 106},~281--287.

\bibitem[Baym(1962)]{Baym62}
Baym, G.
\newblock Self-Consistent Approximations in Many-Body Systems.
\newblock {\em Phys. Rev.} {\bf 1962}, {\em 127},~1391--1401.

\bibitem[Anderson(1961)]{PhysRev.124.41}
Anderson, P.W.
\newblock Localized Magnetic States in Metals.
\newblock {\em Phys. Rev.} {\bf 1961}, {\em 124},~41--53.
\newblock {\url{https://doi.org/10.1103/PhysRev.124.41}}.

\bibitem[Kozik et~al.(2015)Kozik, Ferrero, and Georges]{kozik2015nonexistence}
Kozik, E.; Ferrero, M.; Georges, A.
\newblock Nonexistence of the Luttinger-Ward Functional and Misleading
  Convergence of Skeleton Diagrammatic Series for Hubbard-Like Models.
\newblock {\em Phys. Rev. Lett.} {\bf 2015}, {\em 114},~156402.
\newblock {\url{https://doi.org/10.1103/PhysRevLett.114.156402}}.

\bibitem[Vu\ifmmode \check{c}\else \v{c}\fi{}i\ifmmode \check{c}\else
  \v{c}\fi{}evi\ifmmode~\acute{c}\else \'{c}\fi{} et~al.(2018)Vu\ifmmode
  \check{c}\else \v{c}\fi{}i\ifmmode \check{c}\else
  \v{c}\fi{}evi\ifmmode~\acute{c}\else \'{c}\fi{}, Wentzell, Ferrero, and
  Parcollet]{vuvcivcevic2018practical}
Vu\ifmmode \check{c}\else \v{c}\fi{}i\ifmmode \check{c}\else
  \v{c}\fi{}evi\ifmmode~\acute{c}\else \'{c}\fi{}, J.; Wentzell, N.; Ferrero,
  M.; Parcollet, O.
\newblock Practical consequences of the Luttinger-Ward functional
  multivaluedness for cluster DMFT methods.
\newblock {\em Phys. Rev. B} {\bf 2018}, {\em 97},~125141.
\newblock {\url{https://doi.org/10.1103/PhysRevB.97.125141}}.

\bibitem[Lanczos(1950)]{lanczos1950iteration}
Lanczos, C.
\newblock An iteration method for the solution of the eigenvalue problem of
  linear differential and integral operators.
\newblock {\em Journal of research of the National Bureau of Standards} {\bf
  1950}, {\em 45},~255--282.

\bibitem[Barr et~al.(2012)Barr, Stafford, and Bergfield]{barr2012effective}
Barr, J.D.; Stafford, C.A.; Bergfield, J.P.
\newblock Effective field theory of interacting $\ensuremath{\pi}$ electrons.
\newblock {\em Phys. Rev. B} {\bf 2012}, {\em 86},~115403.
\newblock {\url{https://doi.org/10.1103/PhysRevB.86.115403}}.

\bibitem[Mori(1965)]{mori1965continued}
Mori, H.
\newblock A continued-fraction representation of the time-correlation
  functions.
\newblock {\em Prog. Theor. Phys.} {\bf 1965}, {\em 34},~399--416.

\bibitem[Barford(2005)]{barford2005electronic}
Barford, W.
\newblock {\em Electronic and optical properties of conjugated polymers}; Vol.
  129, OUP Oxford,  2005.

\bibitem[Purcell and Singer(1967)]{purcell1967brief}
Purcell, W.P.; Singer, J.A.
\newblock A brief review and table of semiempirical parameters used in the
  Hueckel molecular orbital method.
\newblock {\em J. Chem. Eng. Data} {\bf 1967}, {\em 12},~235--246.

\bibitem[Ohno(1964)]{ohno1964some}
Ohno, K.
\newblock Some remarks on the Pariser-Parr-Pople method.
\newblock {\em Theor. Chim. Acta} {\bf 1964}, {\em 2},~219--227.

\bibitem[Becke(1993)]{Becke1993}
Becke, A.D.
\newblock Density‐functional thermochemistry. III. The role of exact
  exchange.
\newblock {\em J. Chem. Phys.} {\bf 1993}, {\em 98},~5648--5652.
\newblock {\url{https://doi.org/10.1063/1.464913}}.

\bibitem[Lee et~al.(1988)Lee, Yang, and Parr]{LYP1988}
Lee, C.; Yang, W.; Parr, R.G.
\newblock Development of the Colle–Salvetti correlation-energy formula into a
  functional of the electron density.
\newblock {\em Phys. Rev. B} {\bf 1988}, {\em 37},~785--789.
\newblock {\url{https://doi.org/10.1103/PhysRevB.37.785}}.

\bibitem[Krishnan et~al.(1980)Krishnan, Binkley, Seeger, and
  Pople]{Krishnan1980}
Krishnan, R.; Binkley, J.S.; Seeger, R.; Pople, J.A.
\newblock Self-consistent molecular orbital methods. XX. A basis set for
  correlated wave functions.
\newblock {\em J. Chem. Phys.} {\bf 1980}, {\em 72},~650--654.
\newblock {\url{https://doi.org/10.1063/1.438955}}.

\bibitem[McLean and Chandler(1980)]{McLeanChandler1980}
McLean, A.D.; Chandler, G.S.
\newblock Contracted Gaussian basis sets for molecular calculations. I. Second
  row atoms, Z=11–18.
\newblock {\em J. Chem. Phys.} {\bf 1980}, {\em 72},~5639--5648.
\newblock {\url{https://doi.org/10.1063/1.438980}}.

\bibitem[Neese(2025)]{RN269}
Neese, F.
\newblock Software update: the ORCA program system, version 6.0.
\newblock {\em WIRES Comput. Molec. Sci.} {\bf 2025}, {\em 15},~e70019.
\newblock {\url{https://doi.org/10.1002/wcms.7019}}.

\bibitem[Neese(2003)]{RN33}
Neese, F.
\newblock An improvement of the resolution of the identity approximation for
  the formation of the Coulomb matrix.
\newblock {\em J. Comp. Chem.} {\bf 2003}, {\em 24},~1740--1747.
\newblock {\url{https://doi.org/10.1002/jcc.10318}}.

\bibitem[Grimme et~al.(2010)Grimme, Antony, Ehrlich, and Krieg]{RN75}
Grimme, S.; Antony, J.; Ehrlich, S.; Krieg, H.
\newblock A consistent and accurate ab initio parametrization of density
  functional dispersion correction (DFT-D) for the 94 elements H-Pu.
\newblock {\em J. Chem. Phys.} {\bf 2010}, {\em 132},~154104.
\newblock {\url{https://doi.org/10.1063/1.3382344}}.

\bibitem[Bykov et~al.(2015)Bykov, Petrenko, Izsak, Kossmann, Becker, Valeev,
  and Neese]{RN114}
Bykov, D.; Petrenko, T.; Izsak, R.; Kossmann, S.; Becker, U.; Valeev, E.;
  Neese, F.
\newblock Efficient implementation of the analytic second derivatives of
  Hartree-Fock and hybrid DFT energies: a detailed analysis of different
  approximations.
\newblock {\em Molec. Phys.} {\bf 2015}, {\em 113},~1961--1977.
\newblock {\url{https://doi.org/10.1080/00268976.2015.1025114}}.

\bibitem[Helmich-Paris et~al.(2021)Helmich-Paris, de~Souza, Neese, and
  Izsák]{RN218}
Helmich-Paris, B.; de~Souza, B.; Neese, F.; Izsák, R.
\newblock An improved chain of spheres for exchange algorithm.
\newblock {\em J. Chem. Phys.} {\bf 2021}, {\em 155},~104109.
\newblock {\url{https://doi.org/10.1063/5.0058766}}.

\bibitem[Reddy et~al.(2007)Reddy, Jang, Segalman, and
  Majumdar]{reddy2007thermoelectricity}
Reddy, P.; Jang, S.Y.; Segalman, R.A.; Majumdar, A.
\newblock Thermoelectricity in molecular junctions.
\newblock {\em Science} {\bf 2007}, {\em 315},~1568--1571.

\bibitem[Evangeli et~al.(2013)Evangeli, Gillemot, Leary, Gonzalez,
  Rubio-Bollinger, Lambert, and Agrait]{evangeli2013engineering}
Evangeli, C.; Gillemot, K.; Leary, E.; Gonzalez, M.T.; Rubio-Bollinger, G.;
  Lambert, C.J.; Agrait, N.
\newblock Engineering the thermopower of C60 molecular junctions.
\newblock {\em Nano Lett.} {\bf 2013}, {\em 13},~2141--2145.

\bibitem[Rinc{\'o}n-Garc{\'\i}a et~al.(2016)Rinc{\'o}n-Garc{\'\i}a, Evangeli,
  Rubio-Bollinger, and Agra{\"\i}t]{rincon2016thermopower}
Rinc{\'o}n-Garc{\'\i}a, L.; Evangeli, C.; Rubio-Bollinger, G.; Agra{\"\i}t, N.
\newblock Thermopower measurements in molecular junctions.
\newblock {\em Chem. Soc. Rev.} {\bf 2016}, {\em 45},~4285--4306.

\bibitem[Rankin(2009)]{rankin2009crc}
Rankin, D.W.
\newblock CRC handbook of chemistry and physics, edited by David R. Lide,
  2009.

\bibitem[Solomon et~al.(2008)Solomon, Andrews, Hansen, Goldsmith, Wasielewski,
  Van~Duyne, and Ratner]{solomon2008understanding}
Solomon, G.C.; Andrews, D.Q.; Hansen, T.; Goldsmith, R.H.; Wasielewski, M.R.;
  Van~Duyne, R.P.; Ratner, M.A.
\newblock Understanding quantum interference in coherent molecular conduction.
\newblock {\em J. Chem. Phys.} {\bf 2008}, {\em 129},~054701.

\bibitem[Van~Veen et~al.(2022)Van~Veen, Ornago, Van Der~Zant, and
  El~Abbassi]{van2022benchmark}
Van~Veen, F.H.; Ornago, L.; Van Der~Zant, H.S.; El~Abbassi, M.
\newblock Benchmark study of alkane molecular chains.
\newblock {\em J. Phys. Chem. C} {\bf 2022}, {\em 126},~8801--8806.

\bibitem[Guo et~al.(2011)Guo, Hihath, Diez-Perez, and Tao]{guo2011measurement}
Guo, S.; Hihath, J.; Diez-Perez, I.; Tao, N.
\newblock Measurement and Statistical Analysis of Single-Molecule
  Current--Voltage Characteristics, Transition Voltage Spectroscopy, and
  Tunneling Barrier Height.
\newblock {\em J. Am. Chem. Soc.} {\bf 2011}, {\em 133},~19189--19197.

\bibitem[Li et~al.(2006)Li, He, Hihath, Xu, Lindsay, and
  Tao]{li2006conductance}
Li, X.; He, J.; Hihath, J.; Xu, B.; Lindsay, S.M.; Tao, N.
\newblock Conductance of single alkanedithiols: conduction mechanism and effect
  of molecule- electrode contacts.
\newblock {\em J. Am. Chem. Soc.} {\bf 2006}, {\em 128},~2135--2141.

\bibitem[Li et~al.(2008)Li, Pobelov, Wandlowski, Bagrets, Arnold, and
  Evers]{li2008charge}
Li, C.; Pobelov, I.; Wandlowski, T.; Bagrets, A.; Arnold, A.; Evers, F.
\newblock Charge transport in single Au| alkanedithiol| Au junctions:
  coordination geometries and conformational degrees of freedom.
\newblock {\em J. Am. Chem. Soc.} {\bf 2008}, {\em 130},~318--326.

\bibitem[Haiss et~al.(2009)Haiss, Marti\'in, Leary, Zalinge, Higgins, Bouffier,
  and Nichols]{haiss2009impact}
Haiss, W.; Marti\'in, S.; Leary, E.; Zalinge, H.v.; Higgins, S.J.; Bouffier,
  L.; Nichols, R.J.
\newblock Impact of junction formation method and surface roughness on single
  molecule conductance.
\newblock {\em J. Phys. Chem. C} {\bf 2009}, {\em 113},~5823--5833.

\bibitem[Venkataraman et~al.(2006)Venkataraman, Klare, Tam, Nuckolls,
  Hybertsen, and Steigerwald]{venkataraman2006single}
Venkataraman, L.; Klare, J.E.; Tam, I.W.; Nuckolls, C.; Hybertsen, M.S.;
  Steigerwald, M.L.
\newblock Single-molecule circuits with well-defined molecular conductance.
\newblock {\em Nano Lett.} {\bf 2006}, {\em 6},~458--462.

\end{thebibliography}

\end{document}